\documentclass[letterpaper,12pt]{article}
\pdfoutput=1
\usepackage{jheppub}
\usepackage{epsfig}
\usepackage{amsmath}
\usepackage{mathrsfs}
\usepackage{graphicx}
\usepackage{capt-of}
\usepackage{tikz-cd}
\usepackage{float}
\usepackage{verbatim}
\usepackage{hyperref}
\restylefloat{table}

\definecolor{burgundy}{rgb}{0.5, 0.0, 0.13}

\DeclareSymbolFont{AMSa}{U}{msa}{m}{n}
\DeclareSymbolFont{AMSb}{U}{msb}{m}{n}
\let\Box\relax
\DeclareMathSymbol{\Box}{\mathord}{AMSa}{"03}

\allowdisplaybreaks

\newcommand{\ti}[1]{\textit{#1}}

\newcommand{\IP}[1]{{\langle{#1}\rangle}}

\newcommand{\fL}{{\mathfrak L}}
\newcommand{\fX}{{\mathfrak X}}
\newcommand{\fW}{{\mathfrak W}}

\newcommand{\cL}{{\mathcal L}}
\newcommand{\cA}{{\mathcal A}}
\newcommand{\cM}{{\mathcal M}}

\newcommand{\cN}{{\mathcal N}}

\newcommand{\cS}{{\mathcal S}}
\newcommand{\cO}{{\mathcal O}}

\newcommand{\fg}{{\mathfrak g}}

\newcommand{\fq}{{\mathfrak q}}

\newcommand{\C}{{\mathbb C}}

\newcommand{\bbZ}{{\mathbb Z}}

\newcommand{\bbR}{{\mathbb R}}

\newcommand{\Z}{{\bbZ}}
\newcommand{\R}{{\bbR}}

\newcommand{\abs}[1]{\lvert#1\rvert}

\newcommand{\e}{{\mathrm e}}

\newcommand{\I}{{\mathrm i}}

\newcommand{\tC}{\widetilde C}

\newcommand{\te}{\text{e}}

\newcommand{\lr}[1]{\langle#1\rangle}

\DeclareMathOperator{\Tr}{Tr}
\DeclareMathOperator{\Sk}{Sk}

\DeclareMathOperator{\Spin}{Spin}

\DeclareMathOperator{\SO}{SO}
\DeclareMathOperator{\SU}{SU}

\DeclareMathOperator{\fgl}{{\mathfrak{gl}}}
\DeclareMathOperator{\fsl}{{\mathfrak{sl}}}

\newcommand{\insfigsvg}[3]{

\medskip
\noindent
\begin{minipage}{\linewidth}

\makebox[\linewidth]{\includegraphics[keepaspectratio=true,scale=#2]{figures/#1.pdf}}

\captionof{figure}{#3}

\label{fig:#1}
\end{minipage}
\medskip

}

\title{Commuting Line Defects At $q^N=1$}

\author[a]{Davide Gaiotto,}
\author[b]{Gregory W. Moore,}
\author[c]{Andrew Neitzke,}
\author[b]{and Fei Yan}

\affiliation[a]{Perimeter Institute for Theoretical Physics, Waterloo, Ontario, Canada N2L 2Y5}

\affiliation[b]{NHETC and Department of Physics and Astronomy, Rutgers University, Piscataway NJ 08855, USA}

\affiliation[c]{Department of Mathematics, Yale University, New Haven CT 06511, USA}

\abstract{We explain the physical origin of a curious property of algebras $\cA_\fq$ which encode the rotation-equivariant fusion ring of half-BPS line defects in 
four-dimensional ${\cal N}=2$ supersymmetric quantum field theories. These algebras are a quantization of the algebras of holomorphic functions on the three-dimensional Coulomb branch of the SQFTs, with deformation parameter $\log \fq$. They are known to acquire a large center, canonically isomorphic to the undeformed algebra, whenever $\fq$ is a root of unity. We give a physical explanation of this fact. We also generalize the construction to characterize the action of this center in the $\cA_\fq$-modules associated to three-dimensional ${\cal N}=2$ boundary conditions. Finally, we use dualities to relate this construction to a construction in the Kapustin-Witten twist of four-dimensional ${\cal N}=4$ gauge theory. These considerations give simple physical explanations of certain properties of quantized skein algebras and cluster varieties, and quantum groups, when the deformation parameter is a root of unity. }

\begin{document}

\maketitle
\flushbottom

\section{Introduction}
The quantization of a classical, finite-dimensional phase space is a subtle and ambiguous art \cite{qm1,qm2,qm3}. A typical challenge is to produce a quantization which preserves some desirable properties of the original classical system. 

The study of supersymmetric quantum field theories has offered several novel ways to address such challenges  \cite{Kontsevich:1997vb,Kapustin:2001ij,Kapustin:2005vs,Gukov:2008ve,Nekrasov:2009rc,Nekrasov:2010ka,Witten:2010zr,Yagi:2014toa,Bullimore:2015lsa,Beem:2016cbd,Pestun:2007rz,Alday:2009fs,Drukker:2009id,Gaiotto:2010be,Dimofte:2011py}. The general idea is that protected quantities in certain SQFTs can be organized in the form of an auxiliary one-dimensional quantum mechanical system, whose properties are strongly constrained by the symmetries and dualities of the underlying field theory. 

A typical scenario involves a one-parameter family of supersymmetric systems labelled by the quantization parameter $\hbar$, equipped with an algebra $A_{\hbar}$ of topological local operators which are bound to a line in space-time when $\hbar \neq 0$, but free to move away from the line when $\hbar =0$ \cite{Kontsevich:1997vb,Gukov:2008ve,Yagi:2014toa,Bullimore:2015lsa,Beem:2016cbd}. The product of local operators bound to a line can be noncommutative, but must become commutative when operators can be moved off the line and switch position without colliding. 

In this paper we consider a special situation where the algebra only depends on $\hbar$ through the exponential $q = \te^{2 \pi \I \hbar}$ \cite{Alday:2009fs,Drukker:2009id,Gaiotto:2010be,Dimofte:2011py}. In these situations we find that a subset of the local operators can move off the line for rational values of $\hbar$ other than $0$. This implies that the quantum algebra $A_\hbar$ develops a large center when $\hbar$ approaches these values, with properties controlled by the underlying physical system. 

\medskip

The prototypical example of this phenomenon occurs for the quantum torus algebra, generated by invertible elements $X$ and $P$ with 
\begin{equation}
    X P = q P X \, .
\end{equation}
At $\hbar = 0$, $X$ and $P$ generate the algebra of holomorphic functions on $\mathbb{C}^* \times \mathbb{C}^*$. Whenever $q$ is specialized to an $N$-th root of unity, i.e. $q^N=1$, then $X_N \equiv X^N$ and $P_N \equiv P^N$ become central. These central elements generate a commutative subalgebra, which can be canonically identified with the classical algebra, via the algebra morphism generated by $X_{\mathrm{cl}} \mapsto X_N$ and $P_{\mathrm{cl}} \mapsto P_N$.

This observation may appear rather trivial, but becomes more surprising if we consider a quantum cluster transformation such as 
\begin{equation}
    P \to (1-X) P
\end{equation}
This transformation acts on the center in a remarkably simple manner:
\begin{equation}
    P_N \to (1-X) P \cdots (1-X) P = (1-X) \cdots (1- q^{1-N} X) P_N = (1-X_N) P_N \, , 
\end{equation}
which is identical to the classical limit of the cluster transformation.

Cluster varieties \cite{Fock:2003xxy} are a large class of complex symplectic manifolds equipped with an atlas of  $(\mathbb{C}^*)^d$ coordinate patches related by cluster transformations. They can be quantized systematically by promoting the $(\mathbb{C}^*)^d$ to quantum torus algebras and gluing them together via quantum cluster transformations. 
Parsing through the definitions specialized to $q^N=1$, one derives a very general result: the quantized algebra ${\cal A}_q$ of holomorphic functions on a cluster variety contains a large center if $q$ is a root of unity, which can be canonically identified with the classical algebra ${\cal A}_1$\footnote{There are some subtle sign issues concerning the square root $\fq = q^{\frac12}$ which we will discuss in the main text.}, simply by mapping classical cluster variables to the $N$-th power of the quantum cluster variables. This phenomenon was already pointed out in \cite{Fock:2003xxy}, and has been explored in the physics literature e.g. \cite{Cecotti:2010fi}.

A particularly important type of cluster variety is a character variety, aka moduli space of flat $G$-connections on a punctured Riemann surface $C$. In this case, the quantized algebra ${\cal A}_q$ has a local description on $C$, as a skein algebra of topological line defects in (analytically-continued) Chern-Simons theory. When $\fg$ is of type $A_1$,
the appearance of a large center isomorphic to $\cA_1$ when $q^N = 1$ 
has been proved and used to study
the representation theory, e.g. \cite{Bonahon_2015,Bonahon_2017,korinman2023classical,MR3830231}.
Another particularly important example of a quantized character variety is the 
quantized universal enveloping algebra ${\cal A}_q = U_q(\fg)$; 
there too, the appearance of a large center isomorphic to $\cA_1$ when $q^N = 1$ is a known and important phenomenon, e.g. \cite{MR1103601,MR1124981,BECK199429}. 

\medskip

In this paper we offer a simple and transparent physical explanation for this large-center phenomenon, for 
a particular class of algebras $\cA_q$, namely those which can be defined through the Melvin compactification \cite{Melvin:1963qx,Nekrasov:2003rj,Hellerman:2012zf,Gaiotto:2010be,Braverman:2016wma} of 4d ${\cal N}=2$ gauge theories. In the special case of theories of class $\cS$, our description involves the Melvin compactifications of 6d $(2,0)$ theories, and can be re-expressed in the language of the Kapustin-Witten twist of 4d ${\cal N}=4$ gauge theories \cite{Kapustin:2006pk, Gaiotto:2008sd, Gaiotto:2008ak, Gaiotto:2008sa, Witten:2010cx, Nekrasov:2010ka,Witten:2011zz,Gaiotto:2011nm}. In either description, we explain how judicious configurations of supersymmetric defects both justify and extend the expected root-of-unity phenomena, and offer a variety of potential mathematical conjectures.

\medskip

Our physics perspective is also relevant for a large class of of ``$K_2$-Lagrangian'' submanifolds \cite{Dimofte:2013iv} ${\cal L}$ of cluster varieties, which are essentially built out of cluster transformations and have nice quantizations as modules ${\cal L}_q$ for ${\cal A}_q$. Such manifolds emerge naturally from 3d ${\cal N}=2$ boundary conditions of 4d ${\cal N}=2$ gauge theories \cite{Dimofte:2011py,Beem:2012mb,Dimofte:2013lba}. They also arise in the context of character varieties, as spaces of flat connections and skein modules ${\cal L}_q$ associated to three-dimensional geometries with boundary $C$ \cite{Dimofte:2013lba}. Our physics setup predicts a nice behaviour when $q$ is a root of unity: the center of  ${\cal A}_q$ acting on ${\cal L}_q$ should satisfy the same relations as 
those defining the classical Lagrangian ${\cal L}$. 

\medskip

We conclude this introduction by mentioning just one of the original motivations for this work. The quantization of Coulomb branches of three-dimensional ${\cal N}=4$ SQFTs produces algebras $A_\hbar$ akin to the Weyl algebra (i.e. the universal enveloping algebra of the Heisenberg algebra defined by $[x,p] = \I \hbar $) \cite{Yagi:2014toa,Bullimore:2015lsa,Beem:2016cbd,Braverman:2016pwk,Braverman:2016wma}. In these examples one could scale $\hbar$ to $1$, and hence there cannot be root-of-unity phenomena.  Instead, these quantized algebras can be defined in characteristic $p$ \cite{2005math1247B} and again have a large center which is canonically isomorphic to the classical algebra. This has been used to great effect \cite{2019arXiv190504623W} to study the properties of Coulomb branches, but a physical interpretation of characteristic $p$ calculations is missing.

In practice, the quantization in characteristic $p$ is employed to define interesting sheaves on 3d Coulomb branches. The root-of-unity setting in four dimensions can be used in the same manner to produce sheaves on 4d Coulomb branches, to which we can give a clear physical meaning. The 3d Coulomb branches are limits of 4d Coulomb branches. It would be interesting to compare the sheaves obtained by the two methods. For example, we claim that the 4d sheaves have a natural hyperholomorphic structure, and it would be interesting to know whether this structure persists into the 3d version.

\section{The centers of algebras of line defects at roots of unity} \label{sec:general-N2-centers}

\subsection{Preliminaries on line defects} \label{sec:pre}

The main actors in our construction are half-BPS line defects in a four-dimensional $\cN=2$ supersymmetric quantum field theory we denote as $T$. We refer to \cite{Gaiotto:2010be} for a detailed discussion of the properties of such line defects and a physical derivation of the algebras ${\cal A}_\fq[T]$ associated to these line defects.\footnote{Here $\fq$ is a square root of $q$, whose role will appear momentarily.}

Many of the details of the physical definition of half-BPS line defects are actually unnecessary for this physical derivation. In particular, a physical defect may belong to a large deformation class of defects, which would be identified at the level of the algebra. 

A more modern perspective would employ a holomorphic-topological (HT) twist of the physical theory, i.e. a twist which makes two directions of space-time topological and combines the remaining two in an holomorphic direction. See e.g. \cite{Aganagic:2017tvx,Costello:2020ndc} for a three-dimensional analogue. Supersymmetric line defects map under the twist to topological line defects wrapping one of the topological directions (e.g. ``time'') and can be fused along the transverse topological direction. 

Working equivariantly for the group $\Spin(2)$ of rotations of the holomorphic direction, such topological defects form a (derived) $\C^*$-equivariant monoidal category. The ``collection of line defects'' could be tentatively formalized as the K-theory of such a category. The construction of \cite{2018arXiv180108111C} should be understood as an example of such a construction.

In the following we will write formulae assuming the existence of some countable collection of 
linear generators ${\mathfrak L}_i$ for the K-theory, so that  
the fusion of (classes of) line defects will be a direct sum of (classes of) line defects with coefficients which are K-theory classes of ``Chan-Paton'' vector spaces, which may carry quantum numbers for the symmetries of the problem:
\begin{equation} \label{eq:category-product}
    {\mathfrak L}_i \, {\mathfrak L}_j = \bigoplus_k [V^k_{ij}] \otimes {\mathfrak L}_k \, .
\end{equation}

The language of K-theory is particularly useful in the context of $S^1$ compactifications of the theory. Line defects wrapping $S^1$ appear as local operators in the three-dimensional effective field theory, which only depend on the K-theory class of the original line. Relations such as  \eqref{eq:category-product} capture the product of these 3d local operators, 
while making manifest integrality properties inherited from the underlying 4d setup: the 3d structure constants in this basis will be expressed as characters of the $[V^k_{ij}]$. We will come back to this momentarily in \autoref{sec:ld-local}.

We will not employ the language of twisted SQFT in the rest of this paper, but a translation should be straightforward (and would likely streamline our arguments). Constructions involving six-dimensional $(2,0)$ SCFTs should be analogously recast in terms of their HT twist. 

The Melvin space $M_q$ is the simplest non-trivial three-dimensional ``HT'' geometry, i.e. a manifold equipped with a  transverse holomorphic foliation \cite{Aganagic:2017tvx}. A four-dimensional HT
theory can thus be defined on $\mathbb{R} \times M_q$. This is the context where $\cA_\fq$ will arise.  

In the physical theory, it is natural to consider line operators which are equivariant under the $\Spin(3)$ group of rotations. However, both the HT twist and the operation of fusion are only compatible with equivariance under the $\Spin(2)$ subgroup fixing the topological fusion direction. It would be interesting to find criteria to pin down K-theory classes which include $\Spin(3)$-equivariant defects. We suspect they would generate the collection in an appropriate sense; this conjecture could have interesting mathematical consequences, but we will not study it in this paper.

\subsection{Line defects and local operators} \label{sec:ld-local}

The four-dimensional SQFT $T$ can be compactified on a circle with Ramond spin structure while preserving all supercharges. The resulting effective three-dimensional $\cN=4$ SQFT has a Coulomb branch of vacua.
Upon choosing a parameter $\zeta \in \C^*$, we can identify this Coulomb branch as
a complex symplectic manifold  $\cM_-[T]$, which parameterizes the expectation values of half-BPS line defects wrapping the internal circle.\footnote{The
3d Coulomb branch is a hyperk\"ahler manifold and is thus equipped with a whole $\mathbb{CP}^1$ family of complex structures. In two of these complex structures ($\zeta = 0$ and $\zeta = \infty$), 
the 3d Coulomb branch is a torus fibration over the 4d Coulomb branch, and cannot be parameterized by BPS local operators. In all of the other complex structures ($\zeta \in \C^*$), BPS line defects give holomorphic functions on the 3d Coulomb branch. The choice of $\zeta$ corresponds to the choice of which 1/2-BPS subalgebra of the 3d supersymmetry algebra is preserved by the line defects.
} 
Throughout this paper we will hold $\zeta$ fixed --- say, to $\zeta = 1$ --- and not consider the extra structure coming from varying this choice (with one brief exception, in 
\autoref{sec:3d-perspective}.)
As a complex symplectic manifold, $\cM_-[T]$ has a weak dependence on the radius $R$ of the circle: a change of radius can be absorbed in a rescaling of the dimensionful parameters of $T$ (if any). Thus we will generally hold $R = 1$, except in a few places where we consider the limit of small $R$.

We denote as ${\cal A}_-[T]$ the Poisson algebra generated by the half-BPS line defects wrapping the internal circle. This commutative algebra has integral structure constants (``fusion coefficients'') which capture coarse information about the 
fusion of line defects. If we denote as $L_i$ the local operator obtained by circle-wrapping ${\mathfrak L}_i$, the algebra takes the form
\begin{equation} \label{eq:local-op-product}
    L_i L_j = \sum_k n^k_{ij} L_k \, .
\end{equation}
By taking expectation values we can identify $\cA_-[T]$ with the algebra of global holomorphic functions on $\cM_-[T]$, and under
this identification the product in \eqref{eq:local-op-product} is the usual product of functions.

The circle compactification can be twisted by holonomies for every ``flavour'' symmetry of the theory, i.e. every global symmetry which commutes with the supercharges. 
This gives commutative deformations of $\cA_-[T]$, as the fusion coefficients in the algebra are replaced by characters of fusion spaces, which are Laurent polynomials in appropriate fugacities $\mu$ for the flavour symmetries:
\begin{equation}
    L_i L_j = \sum_{k,f} n^k_{ij}[f] \mu^f L_k
\end{equation}
This is the algebra of functions on ``mass deformations'' of the 3d Coulomb branch, with $\mu$ being combinations of masses and flavour holonomies. Whenever we refer to $\cA_-[T]$, $\cM_-[T]$ etc. below we 
will include the possibility of such deformations. 

All 4d $\cN=2$ theories which we will discuss are equipped with an $SU(2)_R$ R-symmetry.  The $\mathbb{Z}_2$ diagonal combination of the centers $(-1)^F$ and $(-1)^R$ of the centers of the $\Spin(3)$ rotation group and of $SU(2)_R$ commutes with the supercharges and is always available to twist the circle compactification. The result is another 3d Coulomb branch $\cM_+[T]$ and Poisson algebra of functions ${\cal A}_+[T]$.\footnote{Typically, $\cM_\pm[T]$ are isomorphic, though not canonically so. We will see these isomorphisms explicitly in some of the examples later, e.g. in \autoref{sec:pureU1}.} 

In order to ``quantize'' the Poisson algebras ${\cal A}_\pm[T]$, we can further refine the characters of fusion spaces by a fugacity $\fq$ for a certain combination of $\Spin(3)$ and $SU(2)_R$. This is best formulated in terms of a Melvin compactification of the theory, which we now introduce. 

\subsection{The Melvin compactification}\label{sec:setup}

Fix $q \in \C^\times$ with $\abs{q} = 1$. We consider the 4d Euclidean spacetime $$\R \times M_q \, ,$$ where $\R$ is coordinatized by $t$, and $M_q$ is the Melvin space
defined by
\begin{equation}
    M_q = \{ (z,x) \in \C \times \R \} \, / \, [ (z,x) \sim (qz,x+1) ] \, .
\end{equation}

We collect here a few preliminary comments on the topology and geometry of $M_q$, which will be useful in our subsequent discussion.
There is a natural projection $\pi: M_q \to S^1$, defined by $\pi([z,x]) := [x]$. Each fiber of this projection is $\mathbb{C}$. The resulting bundle has structure group $\SO(2)$ and a natural flat connection with holonomy $R_q: z \mapsto qz$. There is also a natural section $\psi: S^1 \to M_q$ defined by $\psi([x]):= [(0,x)]$.
The image of $\psi$ is a closed loop in $M_q$, whose homotopy class generates $\pi_1(M_q) \simeq \mathbb{Z}$. 

When $q$ is a primitive $N^{\mathrm {th}}$ root of unity, the covering map $\mathbb{C}\times \mathbb{R}\to M_q$ generated by the deck transformation $\phi_q: (z,x) \mapsto (qz,x+1)$ factors through 
\begin{equation}\label{eq:factorcover}
\mathbb{C}\times \mathbb{R}\to \mathbb{C}\times \tilde S^1 \to M_q \, ,
\end{equation}
where the first map has deck group $\mathbb{Z}$ generated by $\phi_q^N$, and the second is a smooth $N$-fold covering.
Here $\tilde S^1$ is a circle, best thought of as the quotient of $\mathbb{R}$ by $x \mapsto x + N$. In this case, for every $z_0 \in \mathbb{C}$ there is a map
$\psi_{z_0}: S^1 \to M_q$ defined by $\psi_{z_0}([x]):=[(z_0,Nx)]$. The image of $\psi_{z_0}$ is a loop in $M_q$ which lifts to a closed loop in $\mathbb{C}\times \tilde S^1$ and has a homotopy class which is $N$ times a generator of $\pi_1(M_q)$. Indeed $(\psi_{z_0})_* \pi_*: \pi_1(S^1) \to \pi_1(S^1)$ is multiplication by $N$. For $z_0 \not=0$, the intersection of this loop with the fiber of $\pi$ above any point $[x]$ is the orbit of $z_0$ under the $\mathbb{Z}_N$-action generated by $R_q$. See \autoref{fig:multiwrapping} below. 

$M_q$ has several useful geometric structures. First, since $R_q \in \SO(2)$ the deck transformation $\phi_q$ is an isometry of the Euclidean metric, which thus descends to a flat metric on $M_q$.
Second,
there is an action of ${\mathbb R}$ on $M_q$ by $[(z,x)] \mapsto [(z,x+s)]$, whose orbits give 
a foliation of $M_q$ with $1$-dimensional leaves; this is a transverse holomorphic foliation in the sense of \cite{Aganagic:2017tvx}.
The loops we discussed above are compatible with these geometric structures; indeed,
for every $q$ with $\vert q \vert = 1$, the image of $\psi$ is a compact geodesic, and also a compact leaf of the foliation; similarly, when $q$ is an $N^{\mathrm{th}}$ root of unity, the image of $\psi_{z_0}$, for any $z_0$, is also a compact geodesic and a compact leaf of the HT foliation.\footnote{Compact geodesics in $M_q$ do not exist for other values of $q$; one can draw a line segment from $(z_0,x)$ to $(qz_0,x+1)$ in the covering space, but the resulting closed loop on $M_q$ is not smooth.} 

The manifolds $M_q$ are all diffeomorphic to $\mathbb{C} \times S^1$ because a diffeomorphism of $\mathbb{C}\times \mathbb{R}$ of the form $f:(z,x) \to (z e^{a x}, x)$ conjugates 
any $\phi_q$ to $\phi_1$. But these diffeomorphisms are not isometries, and, more to the point, do not preserve the HT structure. (See section 2 of \cite{Aganagic:2017tvx}.) As HT geometries the $M_q$ are all inequivalent for different $q$.

To define the compactification of our $\cN=2$ theory on $\R \times M_q$,
we need to choose a spin structure. Recall that the set of spin structures on $M_q$ 
is a torsor for $H^1(M_q,\mathbb{Z}_2) \simeq \mathbb{Z}_2$.
Choosing a spin structure is equivalent to choosing
a lift of $R_q = q^{J_3}$ from $\SO(2) \subset \SO(3)$ to the double cover
$\Spin(3) \simeq \SU(2)$. Here $J_3$ is a generator of $SU(2)$ normalized 
so that the eigenvalues of $2J_3$ in the fundamental representation are $\{ \pm 1\}$. 
Such a choice can be written as $\tilde{R}_\fq = (-\fq)^{2 J_3}$ where
$\fq^2 = q$. The two choices of $\fq$ correspond to the two possible spin structures.

When $\fq \neq -1$, the theory on $\R \times M_q$
is not supersymmetric unless we include a flat background $\SU(2)_R$-connection on $M_q$.
This background connection is classified by its holonomy around $S^1$.
We choose this holonomy to be $(-\fq)^{2 I_3}$, where $I_3$ is a generator of $SU(2)_R$, normalized as above. 

With these choices, for generic $q$, the theory on $\R \times M_q$ preserves the $4$ supercharges
with $I_3 + J_3 = 0$.
In the exceptional case $q = 1$ we have 
$M_q = M_1 = \C \times S^1$, so putting our theory on $M_q$ just means
compactifying it on a circle as in
\autoref{sec:ld-local}. 
In that case $\fq = -1$ is the case of trivial $\tilde{R}_\fq$, which 
corresponds to Ramond (periodic) spin structure and no $\SU(2)_R$ twist, while
$\fq=+1$ corresponds to Neveu-Schwarz (antiperiodic) spin structure and a twist
by $-1 \in \SU(2)_R$.
Thus the choice of $\fq = \pm 1$ leads to the algebras $\cA_\pm[T]$ which we discussed
in \autoref{sec:ld-local}.

When $q \neq 1$ is a root of unity we only have $4$ supercharges, just as for generic $q$. Nevertheless, the root of unity case
turns out to have special features, as we discuss in the next section.

\subsection{Line defects in the Melvin background} \label{sec:bulklines}

Now we consider placing supersymmetric line defects in $\R \times M_q$. 
Each line defect will be placed at a point $t \in \R$ and extended along a circle in $M_q$. In order for the line defect to be supersymmetric, this circle must be a closed Euclidean geodesic in $M_q$, i.e. locally a straight line.\footnote{If we used the language of the holomorphic-topological twist, we would say instead that the circle must be a closed leaf of the foliation of $M_q$ which we discussed above. In this example, it happens that the closed geodesics are the same as the closed leaves, so we can use either language.}

When $q \neq 1$ is generic, the only closed Euclidean geodesic in $M_q$ is the one located at $z=0$, i.e. the image of $\psi: S^1 \to M_q$ discussed in the previous section. Thus the space of possible insertion locations for the line defects is $\R_t$. The (K-theory classes of) line defects can be fused along the $\R_t$ direction to give an algebra $\cA_\fq[T]$, whose structure coefficients are Laurent polynomials in $\fq$ with integer coefficients:
\begin{equation}
    L_i L_j = \sum_{k,s} n^k_{ij}[s] \fq^s L_k \, .
\end{equation}
The parameter $\fq$ plays the role of the fugacity for the combination of spin and R-symmetry rotations.
The algebra $\cA_\fq[T]$ is typically not commutative, as wrapped lines cannot exchange their position in $\R_t$ without colliding with each other. 

In the $q \to 1$ limit, we recover the commutative algebras $\cA_\pm[T]$.
From the geometric perspective, this commutativity arises because we now have closed geodesics at any value $z=z_0$, and thus wrapped line defects can be moved away from $z=0$ to exchange their position in $\R_t$ without colliding with each other. 

It is also interesting to consider the behavior \ti{near} $q = 1$.
One geometric way to realize this is as follows. We rescale the metric in $M_q$ by a factor of $R$, so that the geodesic at $z = 0$ now has length $R$. Next we set $q = \e^{2 \pi \I R \hbar}$, 
and take $R \to 0$ while holding $\hbar$ fixed. In that limit we are effectively reduced to a 3-dimensional  ${\cal N}=4$ theory on $\R^3$, in $\Omega$-background \cite{Moore:1997dj,Moore:1998et,Nekrasov:2002qd}, with $\hbar$ playing the role of the $\Omega$-background parameter.
The wrapped line defects reduce to local operators of the ${\cal N}=4$ theory, and their
commutator gives a deformation quantization of a naturally-defined Poisson bracket in $\cA_\pm[T]$ 
(see e.g. \cite{Yagi:2014toa} and Section 6 of \cite{Beem:2018fng}).

Our description in terms of four-dimensional field theory on $M_q$ extends this
deformation quantization,
and shows that the integrality properties of $\cA_\pm$ survive quantization, leading to Laurent polynomials in $\fq$.

\insfigsvg{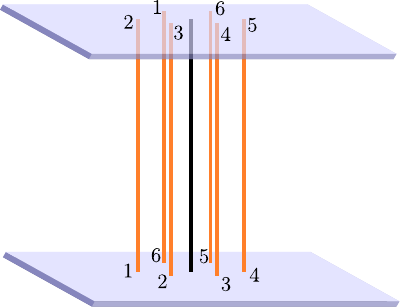}{0.95}{A multiwrapped line defect in $M_q$, in the case $N=6$, $q = \te^{2 \pi \I / 6}$. The black segment is the circle $z = 0$, while the orange segments are pieces of the multiwrapped defect. The two planes are identified by a rotation $z \mapsto q z$. After this identification the
orange segments fit together into a single circle, which is a closed geodesic in $M_q$.}

Now suppose $q$ is a primitive $N$-th root of unity, i.e. $q = \te^{2 \pi \I \frac{m}{N}}$ with $(m,N) = 1$. In this case, as we remarked above, something special happens: we have
again a continuous family of supersymmetric embeddings for a line defect,
given by the loops $\psi_{z_0}$ in $M_q$ for any $z_0 \in \C$.
This is illustrated in \autoref{fig:multiwrapping} for the example of $N=6$.
Thus, any given line defect ${\mathfrak L}$ can be wrapped either on the 
core circle at $z = 0$ to give an element $L_{\mathfrak L}$ of $\cA_\fq$, or can be multiwrapped at a generic $z=z_0$.

Crucially, we can deform the multiwrapped line defect by bringing $z_0 \to 0$. The result will not be $L_{\mathfrak L}$! 
Instead, it will be some other element $L_{\mathfrak L,N}$ in $\cA_\fq$, which is central: it can be deformed back to generic $z_0$ and moved across any other generator in $\cA_\fq$.

Now let us discuss the fusion of the multiwrapped line defects. Each of the geodesics $\psi_{z_0}$ has an open neighborhood of the form $U \times \R \times \tilde{S}^1$, where $U$ is 
a disc in $\R^2$.
Moreover, if we have two multiwrapped line defects on nearby geodesics $\psi_{z_0}$, $\psi_{z'_0}$,
we can find a single neighborhood $U \times \R \times \tilde{S}^1$ containing both of them.
Since 
this neighborhood is isometric to an open subset in $\R^3 \times \tilde{S}^1$, the fusion product of these line defects will be the same as the fusion product of line defects in $\R^3 \times \tilde{S}^1$. 

We need to discuss a detail concerning signs. As we have discussed in \autoref{sec:setup},
there are two possible ways of making the theory supersymmetric in $\R^3 \times \tilde{S}^1$,
labeled by a parameter $\tilde \fq = \pm 1$: around $\tilde{S}^1$ we take $SU(2)_R$ holonomy $- \tilde \fq$ and a spin structure which is Ramond if $\tilde \fq = -1$ and NS if $\tilde \fq = 1$. What is the value 
of $\tilde \fq$ which actually occurs in our setup?
Since $\tilde{S}^1$ is homologous to $N$ times $S^1$,
the $SU(2)_R$ holonomy around $\tilde{S}^1$ is
$(-\fq)^N$; it follows that\footnote{In the language of effective quantum mechanics on the line defect, we might describe the
derivation of \eqref{eq:tildeq-1} as follows. The partition function of the singly-wrapped defect $\fL$ can be written as a supertrace over a defect Hilbert space ${\mathbb V}_\fL$, with a twist by the rotation and $R$-symmetry, of the form
\begin{equation}
\Tr_{{\mathbb V}_\fL} (-1)^{2 J_3} (-\fq)^{2 I_3 + 2 J_3} \e^{-R H} \, . 
\end{equation}
Here the factor $(-1)^{2 J_3}$ is the standard factor for a compactification with the periodic spin structure, and the factor $(-\fq)^{2 I_3 + 2 J_3}$ accounts for the action of the spacetime and $R$-symmetry rotations covering the deck transformation 
$\phi_q: (z,x) \mapsto (qz,x+1)$. The pullback of the periodic (nonbounding) spin structure on the circle to a covering circle is still the periodic spin structure. However,   now we have the deck transformation $\phi_q^N$ instead of $\phi_q$; thus the partition function for the multi-wrapped defect is
\begin{equation}
\Tr_{{\mathbb V}_\fL} (-1)^{2 J_3} ((-\fq)^{2 I_3 + 2 J_3} \e^{-R H})^N = \Tr_{{\mathbb V}_\fL} (-1)^{2 J_3} ((-\fq)^N)^{2 I_3 + 2 J_3} \e^{-N R H} \, ,
\end{equation}
which is related to the singlewrapped one by the replacements
$-\fq \mapsto (-\fq)^N$ and $R \mapsto NR$.
For the fusion algebra we consider, the change of $R$ is irrelevant.
}
\begin{equation} \label{eq:tildeq-1}
-\tilde \fq = (-\fq)^N \, .
\end{equation}
For later convenience we rewrite this
slightly: when $q$ is a primitive $N$-th root of unity, 
$(-\fq)^N = -\fq^{N^2} \in \{\pm 1\}$ both for $N$ even and odd, and thus
\eqref{eq:tildeq-1} becomes
\begin{equation}
\tilde \fq = \fq^{N^2} \, .
\end{equation}
Altogether, then, we conclude that the fusion algebra of the multiwrapped line defects 
is the commutative algebra $\cA_{\pm}$, where the sign $\pm 1 = \tilde \fq = \fq^{N^2}$.

We now make an important assumption, that the K-theory class of the multi-wrapped ${\mathfrak L}$ defect only depends on the K-theory class of the  circle-wrapped defect $L_{\mathfrak L}$.
Under this assumption, we can define a homomorphism of algebras
\begin{equation}
\cA_{\pm} \to \cA_\fq
\end{equation}
by
\begin{equation}\label{eq:IncludeApm} 
    L_{\mathfrak L}[\tilde{\fq} = \pm 1] \mapsto L_{\mathfrak L,N}[\fq] \, .
\end{equation}
We call this the \ti{multiwrapping map}. Its image is central in $\cA_\fq$.

How complicated should the multiwrapping map be in practice? A powerful principle is that the answer is independent of the coupling, and 
thus the calculation can be done in the most convenient duality frame. In particular, for 
Wilson line defects in gauge theories we can get good control by working at weak gauge coupling. This kind of argument has been used to predict that Wilson lines should fuse according to the tensor products of their respective representations:
\begin{equation}
    L_{\fW[R]} L_{\fW[R']} = L_{\fW[R \otimes R']} 
\end{equation}
where $\fW[R]$ is a Wilson line in representation $R$.
In a similar spirit, we conjecture (and verify below in examples) that in $G$ gauge theory we have
\begin{equation} \label{eq:adams}
L_{\fW[R],N} = L_{\fW[\psi^N(R)]} \, ,
\end{equation}
where $\psi^N$ is an Adams operation on the K-theory of the category of $G$-representations, which literally encodes the concept of a multi-wrapped Wilson line:
\begin{equation}
	\Tr_{\psi^N(R)} (g)  = \Tr_R (g^N) \, .
\end{equation}

\subsection{Line defects in the IR and Seiberg-Witten theory}

At generic points $u$ of the 4d Coulomb branch, the theory $T$ admits a low-energy effective description in the form of an abelian gauge theory. Any BPS line defect ${\mathfrak L}$ also admits an effective description in terms of abelian line defects $\mathfrak{X}_\gamma$ with gauge and flavour charges $\gamma$.
The qualifications we made in \autoref{sec:pre} apply here: rather than the full
physical line defects, at best we can discuss K-theory classes in some category of topological line defects in a twist of the theory.\footnote{It would be rather interesting to actually discuss the ``RG flow functor'' mapping the category of UV line defects into a category of IR line defects \cite{2018arXiv180108111C}, perhaps in the same sense as in the web description of categories of 2d TFTs \cite{Gaiotto:2015aoa}.}
With this understood we write the UV-IR map as
\begin{equation}
    \fL \quad \rightsquigarrow \quad \bigoplus_\gamma \left[ V_{\fL, \gamma}\right] \otimes \mathfrak{X}_\gamma
\end{equation}
where $V_{\fL, \gamma}$ are (perhaps virtual) Chan-Paton vector spaces.

The framed BPS degeneracies introduced in \cite{Gaiotto:2010be} agree with
the characters of these Chan-Paton vector spaces,
\begin{equation}\label{eq:FramedBPS2}
\overline{\underline{\Omega}}(\mathfrak L,\gamma,\fq)= {\rm Tr}_{V_{\fL, \gamma}} (-1)^{2J_3} (-\fq)^{2I_3 + 2J_3} \, .
\end{equation}
These framed BPS degeneracies jump at walls of framed marginal stability in the Coulomb branch.
Accordingly, the IR line defects $\fX_\gamma$ also jump, by cluster-like transformations controlled by the BPS spectrum of the theory.

As we did in \autoref{sec:bulklines}, we can
wrap the IR line defects $\fX_\gamma$ around the 
core circle $\psi_0$, thus obtaining generators $X_\gamma := L_{\fX_\gamma}$ of a noncommutative algebra $\cA^{\mathrm{IR}}_\fq$ of local operators on the line.
Moreover, $\cA^{\mathrm{IR}}_\fq$ can be described explicitly: it is the quantum torus \cite{Gaiotto:2010be}
\begin{equation} \label{eq:quantum-torus}
    X_\gamma X_{\gamma'} = \fq^{\langle \gamma, \gamma' \rangle} X_{\gamma+\gamma'} \, .
\end{equation}
The fusion of line defects commutes with the flow from UV to IR; thus this flow gives a
homomorphism $\cA_\fq \to \cA^{\mathrm{IR}}_\fq$.

All of our arguments about the multiwrapping map apply just as well in the IR theory as in the UV theory.
Moreover, since the IR theory is an abelian gauge theory, and any IR line defect
is a Wilson line in some duality frame,
we can determine the multiwrapping map $\cA^{\mathrm{IR}}_{\pm 1} \to \cA^{\mathrm{IR}}_\fq$ completely
using the Adams operations: indeed \eqref{eq:adams} just becomes
\begin{equation} \label{eq:multiwrapping-abelian}
X_{\gamma,N} = X_{N \gamma} \, .
\end{equation}
Now we can make a consistency check: we see directly from the quantum torus relation \eqref{eq:quantum-torus} that $X_{N\gamma}$ is central and
\begin{equation}
    X_{N\gamma} X_{N\gamma'} = \fq^{N^2\langle \gamma, \gamma' \rangle} X_{N(\gamma+\gamma')} \, ,
\end{equation}
so the multiwrapped line defects indeed obey the relations of $\cA^{\mathrm{IR}}_{\pm 1}$,
with the sign $\pm 1 = \fq^{N^2}$ which we derived from the holonomy of the R-symmetry connection in \autoref{sec:bulklines}.

As we have already remarked, 
the local geometry around the multiwrapped defects is precisely the same as for the singly-wrapped
defects. Moreover, the UV-IR map can be computed using only the local geometry around the defect and the constant vevs of Coulomb branch scalars of the 4d theory. It follows that the UV-IR map commutes with the multiwrapping map.
Thus, given a wrapped line defect $L$ (arising as $L = L_\fL$ for some $\fL$), we may compute its multiwrapped image $L_N$
by first flowing to the IR and then applying the simple rule \eqref{eq:multiwrapping-abelian} to each IR summand.
To summarize: 
if in some chamber $L$ has the IR expansion
\begin{equation} \label{eq:ir-expansion}
    L^{\mathrm{IR}} = \sum_\gamma c_\gamma[L,\fq] X_\gamma \, ,
\end{equation}
then the image of $L$ under the multiwrapping map
defined in \autoref{sec:bulklines}
has IR expansion 
\begin{equation} \label{eq:LN-proposed}
    L_N^{\mathrm{IR}} = \sum_\gamma c_\gamma[L,\fq^{N^2}] X_{N\gamma} \, .
\end{equation}
This is the main result of this section: it gives a simple way of computing $L^{\mathrm{IR}}_N$ in terms of $L^{\mathrm{IR}}$.

We should observe that for gauge theories, the IR image of a non-abelian Wilson line is typically {\it not} a direct sum of abelian Wilson lines. The compatibility between our IR and UV predictions for multi-wrapped Wilson lines thus appears to impose strong constraints on the possible framed BPS spectra of non-abelian Wilson lines. We will explore this phenomenon in examples. 

Setting aside the physical discussion for a moment, 
suppose we regard the right side of \eqref{eq:LN-proposed} as the definition of an object 
$L_N^{\mathrm{IR}}$. Then it is easy to see that the composite map
$L[\fq^{N^2}] \mapsto L^{\mathrm{IR}}[\fq^{N^2}] \mapsto L^{\mathrm{IR}}_N$ is a
homomorphism $\cA_{\fq^{N^2}} \to \cA^{\mathrm{IR}}_\fq$,
with central image.
What is not obvious is that $L^{\mathrm{IR}}_N$ so defined
is actually the IR image of an element $L_N \in \cA_\fq$.
Nevertheless, this follows from our physical construction of $L_N$ in terms
of multiwrapping line defects.

It also follows from this construction that
\eqref{eq:LN-proposed} must be consistent with framed wall-crossing. This requires the quantum wall-crossing transformation of $X_{N \gamma}$ to match the classical wall-crossing transformation of $X_\gamma$. This is indeed the case, in a rather non-trivial manner.
Indeed, in general, a quantum wall-crossing transformation multiplies $X_\gamma$ by products of factors of the form
\begin{equation} \label{eq:ks-factor}
    \prod_{s=1}^{\langle \gamma', \gamma \rangle} (1- (- \fq)^m q^{-s}X_{\gamma'})
\end{equation}
or analogous ones depending on certain signs \cite{ks1,Gaiotto:2010be}.\footnote{Here $m$ is related to the spin of ``halo particles'' used in one of the derivations of the wall crossing formula found in \cite{Gaiotto:2010be}. In the wall-crossing formula one takes the product over $m$ in a spin representation. The factors are  raised to integral powers denoted $a_{m,\gamma}$ in \cite{Gaiotto:2010be}.}
If we make the replacement $\gamma \to N \gamma$, the product \eqref{eq:ks-factor} becomes
\begin{equation}
    \prod_{s=1}^{N\langle \gamma', \gamma \rangle} (1- (- \fq)^m q^{-s}X_{\gamma'}) \, ,
\end{equation}
which telescopes nicely to 
\begin{equation}
    (1- (- \fq)^{N m}X_{N \gamma'})^{\langle \gamma', \gamma \rangle} \, ,
\end{equation}
giving the expected classical wall-crossing transformation with sign $\fq^{N^2}$.

We conclude that different choices of IR chamber give equivalent definitions of $L_N$. This means, in particular, that we can try to check our claim that $L_N$ belongs to $\cA_\fq$ by going to a convenient IR chamber. 

The simplest possibility is that in some chamber we have $L = X$, for some $X = X_\gamma$ (i.e., in this chamber the line defect supports only a single framed BPS state). Then we see immediately that 
\begin{equation} \label{eq:one-state-ir}
L_N = L^N \, .
\end{equation}
The relation \eqref{eq:one-state-ir} then holds universally, independent of a choice of chamber.
In particular, it implies that $L_N$ is in $\cA_\fq$ as expected,
since $L$ is.

A more interesting example occurs in $A_1$ theories of class ${\cal S}$.
In these theories, many line defects can be expanded as
\begin{equation} \label{eq:a1-simple-ir}
    L = X + X^{-1}
\end{equation}
in specific chambers, matching their description as Wilson lines in a weakly-coupled $SU(2)$ gauge theory.\footnote{This happens when the 
line defect $\fL$ is represented by a simple closed curve on the internal 
Riemann surface $C$, and the chamber is ``Fenchel-Nielsen'' type in the sense of \cite{Hollands:2013qza}.} 
Then \eqref{eq:LN-proposed} gives $L_N = X^N + X^{-N}$, and thus
\begin{equation} \label{eq:chebyshev-relation}
    L_N = 2 T_N(L/2)
\end{equation}
where $T_N$ are Chebyshev polynomials of the first kind. 
For instance, $L_2 = L^2 - 2$, and $L_3 = L^3 - 3 L$.

It follows from \eqref{eq:chebyshev-relation} that $L_N$ defined
by \eqref{eq:LN-proposed} indeed lies in $\cA_\fq$, as desired. 
Thus we recover a fact proven by different means in \cite{Bonahon_2015}:
the root-of-unity quantization of $A_1$ character varieties
has a large center generated by the elements $L_N = 2 T_N(L/2)$.
We note also that \eqref{eq:chebyshev-relation} matches the expected \eqref{eq:adams}.

This discussion generalizes
to all line defects $L$ which can be realized as UV Wilson lines in some duality frame: one can typically define a weakly-coupled IR chamber where the Wilson lines $L$ are identified with Weyl-invariant polynomials in electric Darboux variables, and their multiwrappings $L_N$ can be identified as other linear combinations of Wilson lines, just as predicted by \eqref{eq:adams}.
We will exhibit some 
explicit examples in $A_1$ and $A_2$ theories in \autoref{sec:SU2Nf4} below.

We find it remarkable that the multiwrapping map $L \mapsto L_N$ encodes properties of line defects in a manner which is independent of the duality frame or IR chamber. It would be tempting
to try to classify line defects according to how multiwrapping acts on them. For instance, we could call a line defect ``pure'' if it has $L_N = L^N$. (We have seen above that this is the case for any line defect which has a single framed BPS state in some chamber, but we stress that the property of being pure is a UV property, independent of any particular chamber.) More generally we could consider, for example, the class of line defects which have $L_N = F_N(L)$ for some polynomial function $F_N$.

\subsection{Line defects on the boundary} \label{sec:line-ops-boundary}

We can also equip the 4d $\cN=2$ theory $T$ with a half-BPS boundary condition $B$. Such a boundary condition supports a collection of BPS line defects, 
which can fuse with each other along the boundary, and also fuse with bulk line defects brought to the boundary.  
We now repeat all of our earlier constructions in the presence of the boundary. 

First, we can consider $T$ on $[0,\infty)_t \times \R^2 \times S^1$, with the boundary condition $B$ at $\{t = 0\} \times \R^2 \times S^1$.
Then wrapping boundary line defects on $S^1$ gives operators which are free to move in the boundary $\R^2$, and thus form a commutative algebra $\cL_\pm$.
The algebra $\cL_\pm$ is also acted on by $\cA_\pm$, since the wrapped line defects in the bulk can fuse with the wrapped line defects on the boundary.

Let us discuss the structure of $\cL_\pm$
as a module over $\cA_\pm$.
Given a point $x \in \cM_\pm$ 
with corresponding maximal ideal $I_x \subset \cA_\pm$, we can consider the quotient algebra $\cL_\pm / I_x \cL_\pm$;
letting $x$ vary this gives a sheaf of
quotient algebras over $\cM_\pm[T]$.
What can we say about this sheaf?
Physically, the boundary condition $B$ descends to a boundary condition for the 3d theory, preserving 2d $(2,2)$ supersymmetry.\footnote{There is some freedom in the choice of which $\cN=(2,2)$ subalgebra is preserved; we choose it in such a way that the bulk line defects with parameter $\zeta$ preserve $2$ of the supersymmetries.}
Such a boundary condition should be supported on some complex Lagrangian submanifold
$\cM_\pm[B]$ of the
3d Coulomb branch $\cM_\pm[T]$, consisting of the 3d vacua compatible with the boundary. We thus expect the sheaf representing $\cL_\pm$ to be supported on $\cM_\pm[B] \subset \cM_\pm[T]$. 

We can now ``quantize'' $\cL_\pm$, by placing $T$ on $[0,\infty)_t \times M_q$,
with the half-BPS boundary condition $B$ at $\{t=0\} \times M_q$.
Boundary line defects wrapping the core circle at $z=0$ in $M_q$
form an $\cA_\fq$-module $\cL_\fq$, with a distinguished vector corresponding to the trivial defect.
With respect to bases given by boundary line defects,
the entries of the matrices giving the action of $\cA_\fq$ in $\cL_\fq$ are again integral Laurent polynomials in $\fq$. 

For $q \neq 1$ the boundary operators have no room to move, and thus $\cL_\fq$ does not have an algebra structure. When $q = 1$ we recover $\cL_\pm$.
 
When $q$ is a root of unity we can also consider multiwrapped line defects
on the boundary. Following the same 
arguments as we used in the bulk, we 
can see that these multiwrapped boundary defects make up a commutative algebra, identified with $\cL_\pm$, again with the sign $\pm 1 = \fq^{N^2}$.
We illustrate single-wrapped and multiwrapped boundary lines in \autoref{fig:bulkbdy}.

\insfigsvg{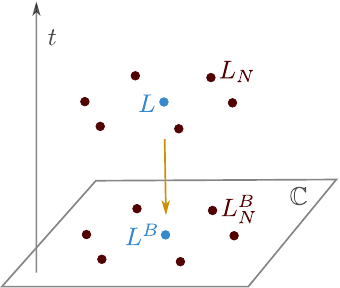}{0.9}{Single-wrapped and multiwrapped lines, both in the bulk ($t > 0$)
and on the boundary ($t = 0$) of $[0,\infty) \times M_q$, in the case $N=6$. We have illustrated the fiber above a single point $[x]\in S^1$.  }

As we have emphasized, $\cL_\fq$ is not an algebra, so 
we cannot literally say that $\cL_\pm$ is ``central'' in $\cL_\fq$. 
Nevertheless, the multiwrapping 
map $\cL_\pm \to \cL_\fq$ plays a role
on the boundary analogous to the role the multiwrapping map $\cA_\pm \to \cA_\fq$ plays in the bulk. Furthermore, we can combine a multi-wrapped boundary line with a standard boundary line at the origin to give an action 
$\cL_\pm \times \cL_\fq \to \cL_\fq$, promoting $\cL_\fq$ to a module for  $\cL_\pm$.

In particular, we can use this map to study the structure of $\cL_\fq$.
Using the bulk multiwrapping map we have
an action of $\cA_\pm$ on $\cL_\fq$,
and the
quotients $\cL_\fq / I_x \cL_\fq$ 
for $x \in \cM_\pm[T]$
are still supported on the Lagrangian subspace $\cM_\pm[B]$. They are modules for the corresponding quotients $\cA_\fq / I_x \cA_\fq$.
Finally, if $\cL_\pm$ is not generated by $\cA_\pm$ acting on the identity, we can further decompose these modules
under the action of the commutative algebra $\cL_\pm$, to get smaller sheaves over the spectrum of $\cL_\pm$.

\subsection{An orbifold defect in 3d} \label{sec:3d-perspective}

In this section we briefly discuss another point of view on our construction.
Assume that $q$ is a primitive $N^{\mathrm{th}}$ root of unity. We can dimensionally reduce 
along the fibration 
\begin{equation}\label{eq:CircleBundle}
    \pi_2: M_q \to \mathbb{C}/\mathbb{Z}_N
\end{equation} 
defined by $\pi_2: [(z,x)] \mapsto [z]$.\footnote{This is a circle bundle over an orbifold, or more properly, a circle bundle over a quotient groupoid (i.e. a quotient stack).}
The reduced theory is thus a 3d theory on a supersymmetric orbifold $\C / \mathbb{Z}_N \times \R$.
We can think of this as the 3d theory in the presence of a codimension-2 supersymmetric defect,
supported at the orbifold line. Passing to an appropriate topological twist of the 3d theory, the 
orbifold should be invisible and we should be able to treat this as a standard line defect. 

We can then move to the 3d Coulomb branch, where the 3d theory has an effective description as an IR-free non-linear sigma-model with target $\cM_\pm[T]$. In this setup a supersymmetric codimension-2 defect should admit an effective description in terms of some auxiliary finite-dimensional quantum-mechanical system coupled to the sigma model, i.e. a hyperholomorphic sheaf 
$V_\fq$ on $\cM_\pm[T]$ which is a finite-rank vector bundle at least over a Zariski open subset of $\cM_\pm[T]$ \cite{Thompson:2000pw}.
Considering UV line defects at $z = 0$ as local operators on the orbifold defect would give a map from $\cA_\fq$ to the space of global sections of ${\mathrm{End}}(V_\fq)$, compatible with the identification of the multiwrapped subalgebra
$\cA_\pm \subset \cA_\fq$ with functions on $\cM_\pm[T]$.

We will now assume that this map is actually an isomorphism. This is a non-trivial assumption: it means that operators which are different in the UV will act differently on the IR data, and every possible operator in the IR description is the image of some UV operator. We will not try to justify this assumption. It is
analogous to the observation that $\cA_\pm$ is exactly the algebra of holomorphic functions on $\cM_\pm$, rather than just mapping into it; that appears to be true in practice, but we are not aware of a proof of it.\footnote{A further layer of subtlety arises from the fact that $\cM_\pm$
may not be affine. Indeed, the spectrum of $\cA_\pm$ often has conical singularities, and if these are resolved by real mass parameters alone, then $\cM_\pm$ is not affine. Generic complex masses, though, should deform the geometry and remove the problem.}

At any rate, under this assumption, we could glean information about $\cA_\fq$  by considering the sheaf of quotient algebras\footnote{Note that $I_x$ is central, so $I_x \cA_\fq=\cA_\fq I_x$ is a two-sided ideal and hence the quotient is an algebra.}
$\cA_\fq / I_x \cA_\fq$
over $\cM_\pm[T]$, where $I_x \subset \cA_\pm$ are the maximal ideals associated to points $x \in \cM_\pm[T]$: for sufficiently general $x$, these quotients should generically be finite-dimensional matrix algebras, the fibers of ${\mathrm{End}}(V_\fq)$.

In various concrete cases, the statement that the fibers $\cA_\fq / I_x \cA_\fq$ are matrix algebras for sufficiently
general $x \in \cM_\pm$ is known in the mathematics
literature --- see e.g. \cite{Bonahon_2015,Bonahon_2017,bonahon2015representations,karuo2022azumaya} for
skein algebras, \cite{MR1124981,BECK199429} for quantized universal enveloping algebras, and
\cite{Fock:2003xxy} for cluster varieties.
One would like to understand more precisely what ``sufficiently general $x$'' 
should mean; for the $\fsl(2)$ skein algebras this question is answered precisely in \cite{karuo2022azumaya},
and it would be interesting to understand the answer in physical terms.

One new statement we
are making here is that the sheaf of algebras $\cA_\fq$ and the sheaf of 
representations $V_\fq$ carry natural hyperholomorphic structures over
the hyperk\"ahler space $\cM_\pm[T]$.
One could go further and attempt to construct these hyperholomorphic structures
concretely.
This can be done explicitly in the simple case where $T$ is a pure 
abelian $\cN=2$ gauge theory. More generally, one should be able to start with the IR abelian
gauge theory and improve it by corrections coming from BPS particles to give the
desired hyperholomorphic connections; this would involve an extension of the 
TBA-type integral equation methods of \cite{Gaiotto:2008cd,hhlb,Gaiotto:2011tf}. 
It would be very interesting to develop this extension in detail.\footnote{The relevant quantum dilogarithm identities have appeared in a related context in \cite{Alexandrov:2015xir}, and a closely related construction of holomorphic bundles over cluster varieties appears in upcoming work of Fock-Goncharov \cite{FGtoappear}.}

\section{Examples}\label{sec:examples}
\subsection{Pure \texorpdfstring{$U(1)$}{U(1)} gauge theory}\label{sec:pureU1}

Let us consider the simplest example, the pure 4d $\cN=2$ gauge theory
with gauge group $U(1)$.
In this theory there are Wilson-'t Hooft line defects ${\mathfrak L}_{\gamma}$ labeled by electromagnetic charge $\gamma = (e,m) \in \Z^2$,with Dirac-Schwinger-Zwanziger pairing $\IP{\gamma,\gamma'} = e m'-m e'$.

We denote the single-wrapped Wilson line with unit positive charge as $L_{1,0} \equiv X$, and the single-wrapped 't Hooft line with unit positive charge as $L_{0,1} \equiv P$. For generic $q\neq 1$, we have the relation in $\cA_\fq$
\begin{equation}
X P = q P X \, ,
\end{equation}
defining the quantum torus algebra. More generally, 
\begin{equation} \label{eq:q-relation-U1}
L_{e,m} L_{e',m'} = \fq^{e m'-m e'} L_{e + e', m + m'} \, .
\end{equation}
Notice that the algebras $\cA_\fq$ and $\cA_{-\fq}$ can be identified by the transformation $L_{e,m} \to (-1)^{e m} L_{e,m}$. The identification is not canonical: it depends on a choice of electric-magnetic duality frame. 
\textbf{}
When $q$ is a primitive $N$-th root of unity, the quantum torus has a large center, which is
precisely spanned by the $L_{\gamma,N} = L_{N \gamma}$, as expected. As discussed in the previous section, the algebra of the $L_{N \gamma}$ can be canonically identified with the commutative algebra $\cA_{\fq^{N^2}}$. 

We can localize the algebra $\cA_\fq$ by setting $X^N = x$, $P^N = p$ for some $x, p \in \C^\times$. The quotient of $\cA_\fq$ by these relations is an algebra of dimension $N^2$. Moreover, for any
given $(x,p)$ we can represent $X$ and $P$ by 
clock and shift operators; this identifies the quotient as the algebra of $N \times N$ complex matrices, in keeping with our discussion in \autoref{sec:3d-perspective}.

Next, we can introduce some half-BPS boundary conditions. 

\subsubsection{The pure Dirichlet / Neumann boundary condition}

 The simplest examples are the Dirichlet and pure 
 Neumann boundary conditions. As they are related by electric-magnetic duality, we only really need to describe one of them. 

At a Dirichlet boundary, the $U(1)$ gauge symmetry becomes a global symmetry. As we compactify this boundary condition on the circle, we get a family of Dirichlet boundary conditions $D_\mu$, labeled by the background holonomy $\mu \in \C^\times$.

The Wilson line brought to the boundary becomes a flavour Wilson line, and thus acting on the boundary we have $X=\mu$. The left ideal
generated by $X-\mu$ determines the left $\cA_\fq$-module $\cL_\fq[D_\mu] = \cA_\fq / \cA_\fq (X-\mu)$. This module is free on the basis $\{ P^r : r\in \Z\}$.
 
Now suppose $q$ is a primitive $N$-th root of unity. On the boundary we have the relation $X_N = \mu^N$, so the commutative algebra $\cL_{\pm}[D_\mu] \subset \cL_\fq[D_\mu]$ of multiwrapped operators on the boundary is freely generated by $P_N = P^N$ alone.

We can localize the algebra $\cA_\fq$ by setting $X^N = x$, $P^N = p$. We find that the 
$\cA_\fq$-module $\cL_\fq[D_\mu]$ is supported on the Lagrangian $x = \mu^N$, as expected. The 
localized module is $N$-dimensional and gives the above-mentioned clock-and-shift representation for the localized algebra.

So far we discussed pure Dirichlet boundary conditions. In the case of pure Neumann boundary conditions all the same discussion applies, with the roles of $X$ and $P$ exchanged.
 
 \subsubsection{Neumann boundary conditions enriched by a chiral multiplet}
 
We can also consider a simple type of enriched Neumann boundary condition: a 3d $\cN=2$ chiral multiplet coupled to the restriction 
of the bulk $U(1)$ gauge field. Let $a \in \Z$ be the
charge of this boundary chiral multiplet.
 
 For $a=1$, line defects brought to the boundary obey the {\it tetrahedron relation} (see \cite{Dimofte:2011ju} equation (7.5)): 
 \begin{equation} \label{eq:tetrahedron-relation}
 X+P=1.
 \end{equation}
The module $\cL_\fq$ is the quotient of $\cA_\fq$ by the left ideal generated by this relation. 
  
 When $q$ is a primitive $N$-th root of unity, our discussion predicts that the multiwrapped lines $X_N$ and $P_N$ should obey the relation defining $\cL_\pm$,
 which is again \eqref{eq:tetrahedron-relation}. We
 can check this directly:
 \begin{equation}
 \begin{split}
 P_N &= P^N = (1-X) P^{N-1} \\
 &= P^{N-1}(1-q^{N-1}X)\\
 &= P^{N-2}(1-q^{N-2}X)(1-q^{N-1}X)
\\ 
 &=(1-X) \cdots (1-q^{N-2}X)(1-q^{N-1}X)\\
 &=1+(-1)^{N}q^{\frac{N(N-1)}{2}}X^N = 1- X_N ~~\text{for} ~~ q=\text{e}^{2\pi\text{i}/N}
 \end{split}
 \end{equation}
 as desired.

The case of general $a$ follows easily from the $a=1$ case by redefining the charge lattice. 
Starting from
\begin{equation}
    P^a = P^{a-1} (1-X) = (1-q^{a-1} X) P^{a-1} = \cdots = (1-X)(1-q X) \cdots (1-q^{a-1} X)
\end{equation}
and replacing $P^a \to P$, $X \to X^a$ (which follows from the relation between the charge lattice of the charge $a$ chiral multiplet and the theory with $a=1$) we obtain the correct generalization
of \eqref{eq:tetrahedron-relation}:
\begin{equation}\label{eq:P-charge-a}
    P = (1-X^a)(1-q X^a) \cdots (1-q^{a-1} X^a) \, .
\end{equation}
Similarly, taking the $a^{\mathrm{th}}$ power of the charge-$1$ relation
\begin{equation}
    P_N^a = (1-X_N)^a \, ,
\end{equation}
and substituting $P_N^a \to P_N$, $X_N \to X_N^a$, we obtain the desired relation
\begin{equation}
    P_N = (1-X_N^a)^a \, .
\end{equation}
(Note that a direct verification starting from \eqref{eq:P-charge-a} is slightly involved.)

\subsection{The \texorpdfstring{$U(1)$}{U(1)} theory with one hypermultiplet} \label{sec:u1-one-hyper}

Now let us briefly discuss the 4d $\cN=2$
theory with gauge group $U(1)$ coupled 
to a hypermultiplet with charge $1$.
In this case the algebra $\cA_\fq$
is slightly more interesting; it is
generated by lines $X$, $Y$, $Z^{\pm 1}$
with relations
\begin{align}\label{eqn:u1onehyper}
    X Y &= 1 + \fq^{-1} Z, \\
    Y X &= 1 + \fq Z, \\
    Y Z &= \fq^2 Z Y, \\
    X Z &= \fq^{-2} Z X.
\end{align}
Here $X$ and $Y$ are 't Hooft loops with
magnetic charge $+1$ and $-1$ respectively, 
and $Z^n$ is the Wilson loop with electric charge $n$.
These equations can be deduced from the wall-crossing formulae in \cite{Gaiotto:2010be}
and the bulk BPS spectrum of the theory.
In particular, each of $X$, $Y$, $Z^{\pm 1}$ has a chamber in which it supports only one
framed BPS state. We can rewrite \eqref{eqn:u1onehyper} as wall-crossing equations on the overlap of chambers. But these overlaps are Zariski open, so the relations hold universally. 
It follows that the multiwrapping map 
is $X_N = X^N$, $Y_N = Y^N$, $Z_N^{\pm 1} = Z^{\pm N}$.

When $q$ is a primitive $N$-th root of unity, we can check
directly that $X^N$, $Y^N$, $Z^{\pm N}$ are central.
Moreover, we check
\begin{align}
    X^N Y^N = X^{N-1} (1 + \fq^{-1} Z) Y^{N-1} &= (1 + \fq Z) X^{N-1} Y^{N-1} \\
    &= (1 + \fq Z) (1 + \fq^3 Z) X^{N-2} Y^{N-2} \\
    &= (1 + \fq Z) (1 + \fq^3 Z) \cdots (1 + \fq^{2N-1} Z) \\
    &= 1 \pm Z^N
\end{align}
with the sign $\pm 1 = \fq^{N^2}$
as usual.
This confirms that the generators $X^N$, $Y^N$, $Z^N$ obey the classical
relations in $\cA_{\pm 1}$, i.e. the
multiwrapping map embeds $\cA_{\pm 1} \to \cA_\fq$, as we expect.

We can also give a geometric representation of the line operators in this theory, using its class $\cS$ realization as the compactification of the 6d $(2,0)$ theory of type $\fsl(2)$ on a sphere with a single irregular singularity. There are four Stokes directions and four anti-Stokes directions associated with this irregular singularity. We illustrate this in \autoref{fig:u1onehyper}, where we cut out an infinitesimal disc around the irregular singularity. Each Stokes direction carries a labeling by an ordered pair of integers, which keep track of the sheets of a double cover of the boundary circle.\footnote{If we think of the irregular singularity as associated to a connection with a pole, the sheets of this covering would be labeling the eigenvalues of the connection form near the pole.}
The 't Hooft lines $X$ and $Y$ correspond to open paths connecting two anti-Stokes directions which are not adjacent to each other; each end of the open paths carries a label, specifying one of the sheets of the double cover.\footnote{This kind of skeins on manifolds with boundary, carrying extra labels at the boundary, are sometimes referred to as \ti{stated} skeins; see \cite{MR4493620}.} Finally, the Wilson lines $Z^{\pm 1}$ are represented by loops traveling around the boundary of the cut-out disc, also carrying a sheet label.

\insfigsvg{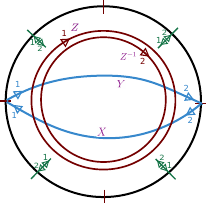}{1.5}{Here we show $\mathbb{CP}^1$ minus an infinitesimal disc around an irregular singularity placed at infinity. There are four Stokes directions at the boundary of the disc labeled by green markers, as well as four anti-Stokes directions labeled by red markers. The numbered arrows indicate the canonical orientations on the double cover of $\mathbb{CP}^1$ minus the disc. The 't Hooft lines $X$ and $Y$ (in blue) correspond to open paths connecting two non-adjacent anti-Stokes directions, while the Wilson lines $Z^\pm$ (in red) correspond to loops around the boundary of the disc.}

The relations obeyed by the line operators now follow from $\fsl(2)$ skein relations on the disc.  Rather than formulating and checking these skein relations directly, it is convenient to use the UV-IR map to an abelian skein algebra associated with a branched double cover of the disc; we can compute this map using the methods of \cite{Neitzke:2020jik}.\footnote{It seems likely that the UV-IR map in this case is equivalent to the stated quantum trace of \cite{MR4431131}; this would be interesting to verify.} For example, the verification of \eqref{eqn:u1onehyper} is illustrated in \autoref{fig:u1onehyperlifts}.

\insfigsvg{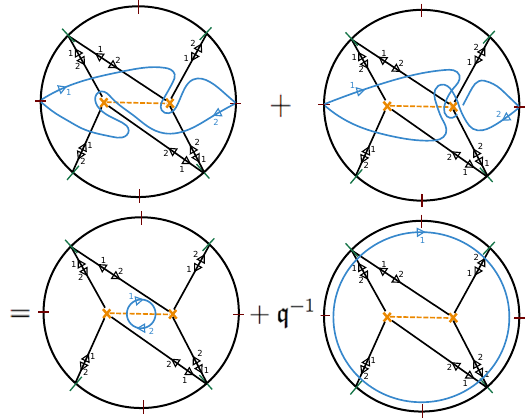}{1.1}{Here we illustrate the two lifts of $XY$ in the abelian skein algebra, with their simplification using the abelian skein relations.}

\subsection{Class $\cS$ theories on the four-punctured sphere}\label{sec:SU2Nf4}

Next we discuss some more elaborate $\cN=2$ theories.
We will use class $\cS$ descriptions of the theories as a convenient tool for computation
\cite{Klemm:1996bj,Witten:1997sc,Gaiotto:2009we,Gaiotto:2009hg}.

\begin{figure}[htbp] \centering \includegraphics[scale=0.16]{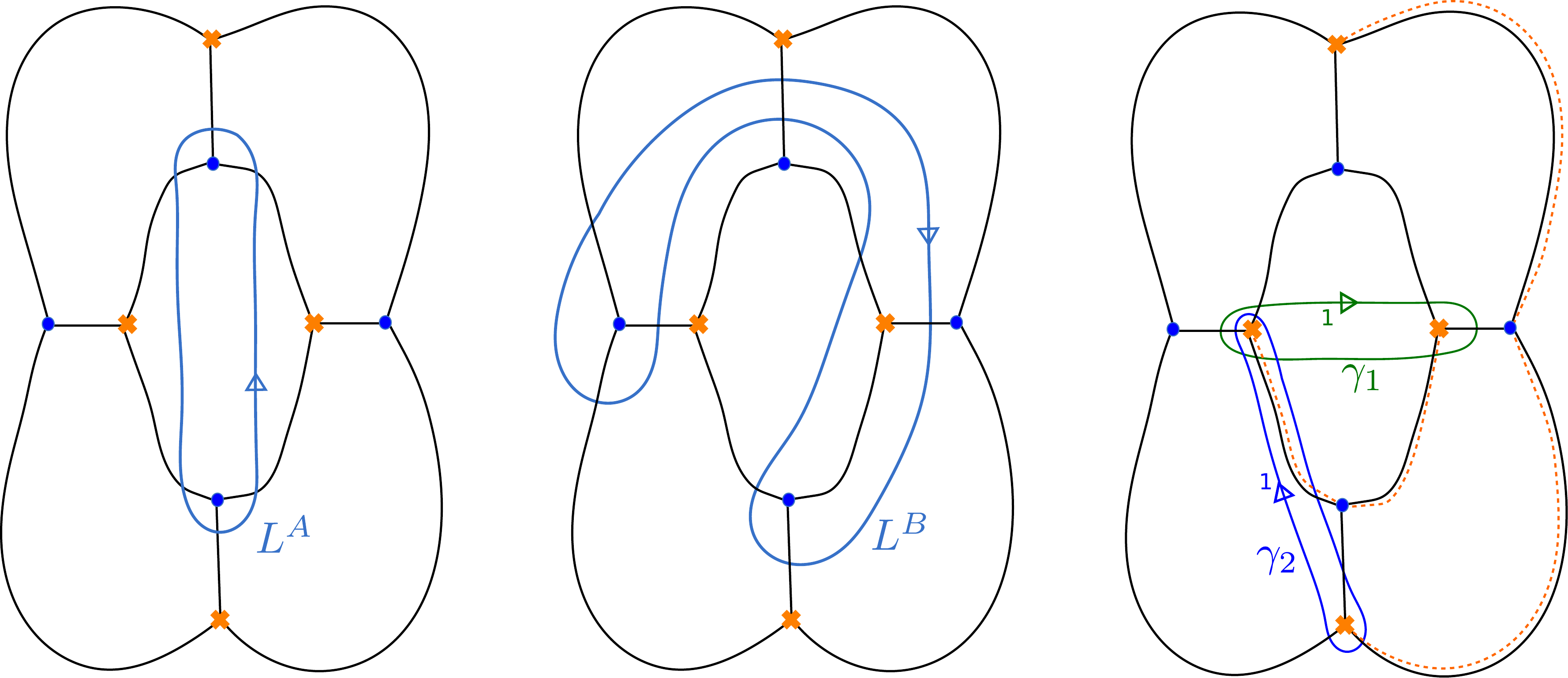} \caption{Left, center: two $1/2$-BPS UV line defects in the $SU(2)$ $N_f=4$ theory, represented as loops on the four-punctured sphere. Right: the IR electromagnetic charges $\gamma_1$, $\gamma_2$, represented as cycles on the Seiberg-Witten curve. The blue dots indicate
the four punctures;
orange crosses indicate the branch points
of the Seiberg-Witten curve \eqref{eq:sw-a1-case}; black arcs show the spectral network associated to the Seiberg-Witten curve.} \label{fig:tetrahedronLD1LD2} \end{figure}

We begin with the $SU(2)$ theory
with $N_f=4$, realized as a class $\cS$
theory of type $A_1$
on the sphere with four regular punctures.
In this theory, BPS 
line defects correspond to isotopy classes
of simple closed curve on the
four-punctured sphere \cite{Drukker:2009id,Gaiotto:2010be}.
We consider the two line defects $L^{A,B}$ shown in Figure \ref{fig:tetrahedronLD1LD2}. 
We will study their IR expansions at a 
point in the strong coupling region, where the framed BPS spectrum is especially simple. The corresponding Seiberg-Witten curve is 
\begin{equation} \label{eq:sw-a1-case}
\tC = \{\lambda: \lambda^2+\phi_2 =0\}\subset T^*C,\quad
\phi_2=-\frac{z^4+2z^2-1}{2(z^4-1)^2} dz^2.
\end{equation}
We express the electromagnetic charges relative to a basis $\{\gamma_1,\gamma_2\}$ with $\lr{\gamma_1,\gamma_2}=-1$,
also shown in the figure.
To shorten expressions we will suppress flavour charges, keeping only the electromagnetic charge.\footnote{This truncation does not affect the commutation relations between the $X_\gamma$, as the flavour charges are in the kernel of the DSZ pairing.} 

The Wilson line $L^A$ has the following IR expansion \cite{Neitzke:2020jik}:\footnote{Here and below, when we refer to \cite{Neitzke:2020jik}, we should note that
$\fq_{\text{here}}=-q_{\text{there}}$.}
\begin{equation}
L^A=X_{(1,0)}+X_{(-1,0)}+X_{(-1,2)}+2X_{(0,1)}+2X_{(-1,1)}.
\end{equation}
Given this expression we can directly compute the Chebyshev
polynomial $T_N(L^A)$ at $q=\text{e}^{2\pi\text{i}/N}$.
Using Mathematica, we verified for $N \le 5$
that it does have the expected expression, obtained by replacing
each $\gamma \to N \gamma$:
\begin{equation}
T_N(L^A)=X_{(N,0)}+X_{(-N,0)}+X_{(-N,2N)}+2X_{(0,N)}+2X_{(-N,N)}.
\end{equation}
The line defect $L^B$ has a more complicated IR expansion, which can
be obtained e.g. using the methods of \cite{Neitzke:2020jik,Bonahon_2015}:
\begin{equation} \label{eq:lb-expansion}
\begin{split}
L^B =& \ X_{(-1,1)}+2X_{(-1,2)}+X_{(-1,3)}+6X_{(0,1)}+2(\fq+\fq^{-1})X_{(0,2)}+X_{(1,-1)}\\&+6X_{(1,0)}+(4+\fq^2+\fq^{-2})X_{(1,1)}+2X_{(2,-1)}+2(\fq+\fq^{-1})X_{(2,0)}+X_{(3,-1)}.
\end{split}
\end{equation}
Again we verified that $T_N(L^B)$ at $q=\text{e}^{2\pi\text{i}/N}$ looks as expected (for $N=2,3$): we replace each $\gamma \to N \gamma$ and also replace $\fq \to \fq^{N^2} = \pm 1$,
\begin{equation}
\begin{split}
T_N(L^B)=&\ X_{(-N,N)}+2X_{(-N,2N)}+X_{(-N,3N)}+6X_{(0,N)} \pm 4X_{(0,2N)}+X_{(N,-N)}\\
&+6X_{(N,0)}+6X_{(N,N)}+2X_{(2N,-N)} \pm 4X_{(2N,0)}+X_{(3N,-N)}
\end{split}
\end{equation}
All this matches 
with the expectations from the previous sections, and with the
theorems of \cite{bonahon2015representations}.
We remark that this provides a nice check on the IR expansions of line defects; if one modifies the expansion \eqref{eq:lb-expansion}, say by changing one of the coefficients, then this match typically does not work out anymore. 

Now let us move on to the case of an $A_2$ theory
of class $\cS$,
again on the four-punctured sphere, with all four punctures ``full'' in the sense of 
\cite{Gaiotto:2009we}.
This theory is an $SU(3)$ gauging of two copies of the Minahan-Nemeschansky $E_6$ theory \cite{Gaiotto:2009we}.

The expectations in the $A_2$ case are slightly more complicated than for $A_1$.
If $L$ is a fundamental Wilson line in a weak
coupling region, and $\overline{L}$
the corresponding antifundamental Wilson line,
then instead of \eqref{eq:a1-simple-ir} we will have
IR expansions of the form
\begin{equation}
    L = X + Y + (XY)^{-1} \, , \quad \overline{L} = X^{-1} + Y^{-1} + XY \,  ,
\end{equation}
where $X$ and $Y$ commute.
Similarly
\begin{equation}
    L_N = X^N + Y^N + (XY)^{-N}, \quad \overline{L}_N = X^{-N} + Y^{-N} + (XY)^N \, .
\end{equation}
Using these expansions we can write $L_N$
as a polynomial in $L$ and $\overline{L}$,
the higher-rank generalization of the 
Chebyshev polynomials which appeared in the 
$A_1$ case.
For instance,
\begin{equation} \label{eq:higher-chebyshev-examples}
    L_2 = L^2 - 2 \overline{L} \, , \qquad L_3 = L^3 - 3 L \overline{L} + 3 \, .
\end{equation}
As before, although we used a weak-coupling
region to derive these relations, it follows that they
hold at any point of the Coulomb branch.

At some points of the Coulomb branch, we can
verify \eqref{eq:higher-chebyshev-examples} directly
using the methods of
\cite{Neitzke:2021gxr}.
Namely, consider a point of the Coulomb branch where
the Seiberg-Witten curve is
\begin{equation}
\tC = \{\lambda: \lambda^3+\phi_2\lambda +\phi_3 =0\}\subset T^*C,
\end{equation}
where $\phi_3$ is very small and $\phi_2$ 
is as in \eqref{eq:sw-a1-case} above.
In this case the IR expansions of the line
defects are quite long; even for $L^A$
we already have $48$ terms $X_\gamma$,
with each $\gamma$ a vector of length $8$
(since in this theory the electromagnetic charge
lattice has rank $8$).
Still, with computer assistance we verified
that the IR expansions of $L^A$ and $\overline{L^A}$
indeed obey the expected relations when $q$ is a primitive $N$-th
root of unity, for $N=2$ or $N=3$.
For instance,
evaluating 
$L^A_2 = (L^A)^2 - 2 \overline{L^A}$ for $q = -1$, 
one obtains again 48 terms, and these 48 terms match 
what one obtains by starting with the IR expansion of $L^A$,
setting $\fq = 1$, and replacing $X_\gamma \to X_{2 \gamma}$
in each term.

The arXiv version of this paper includes an ancillary
Mathematica file which performs these verifications
(including the flavour charges, and also including a slight
additional enhancement to treat the theories of type $\fgl(2)$ and $\fgl(3)$
rather than $A_1$ and $A_2$.)

\subsection{Coulomb branch algebras and quantum groups}

In this section we discuss several classes of quantized Coulomb branch algebras which are related to quantum groups, and the corresponding root-of-unity phenomena.

\subsubsection{Quantum groups and class \texorpdfstring{$\cS$}{S}}

First we consider theories of class $\cS$ \cite{Klemm:1996bj,Witten:1997sc,Gaiotto:2009we,Gaiotto:2009hg}. Recall that these theories are obtained by twisted compactification of 6d $(2,0)$ theories, labelled by an ADE Lie algebra $\fg$, on a Riemann surface $C$ possibly decorated with punctures and/or twist lines. The 3d Coulomb branch of a class $\cS$ theory is a {\it character variety}, i.e. a moduli space of flat $\fg$-connections on $C$.\footnote{There are some important topological subtleties here, relating the choice of global form of the gauge group in the class $\cS$ theory to choices of global form of the gauge group in the space of flat connections. We will ignore these subtleties here.}

The relation to quantum groups arises when we
consider class $\cS$ theories in the case of a surface $C$ with a rank 1 irregular puncture (and any number of other punctures, regular or irregular). 
By a ``rank $1$ irregular puncture'' we mean a puncture where the flat $\fg$-connection has a second-order pole
with semisimple residue.
The Stokes data at such a puncture 
consists of matrices $S^\pm$ in opposite Borel subgroups of $G$, together with a formal monodromy matrix 
$W$ in the Cartan subgroup $H$. The total monodromy around the puncture is $S^+ W S^-$. 

To be precise, the theories we need
to consider are not exactly the usual class $\cS$ theories.
In a class $\cS$ theory with a rank
$1$ irregular puncture, 
there is a global $H$ symmetry;
the formal monodromy $W$ is fixed in terms of mass parameters for that symmetry, and the matrix elements of $S^\pm$ are not
well-defined functions on the 3d Coulomb branch: rather, they are well defined 
only up to the conjugation action of $H$.
We modify the class $\cS$ theory by gauging this global $H$ symmetry. This enlarges the 3d Coulomb branch: it makes $W$ into a function on the Coulomb branch rather
than a fixed mass parameter, and removes the identification under conjugation by $H$. The matrix elements of $S^\pm$, along with $W$, are well defined functions on the enlarged 3d Coulomb branch; thus they give elements of $\cA_1$.

The matrices $S^\pm$ and $W$ associated with a rank $1$ irregular puncture have another interpretation: they parameterize the ``dual Poisson-Lie group'' $G^*$, as originally observed in  \cite{boalchpoissonlie}.\footnote{More precisely, there is an issue of finite coverings here, which we are suppressing in the main text. Given 
opposite Borel subgroups $B_\pm$ of $G$ with $\pi:B_\pm \to H$ the standard
projection, one can 
define $G^*$ as $\{(g_-, g_+) \in B_- \times B_+ \mid \pi(g_-) \pi(g_+) =1$\}. 
The Stokes data $S^{\pm}$, $W$ are related to $g_\pm$ 
by $g_+ = S^+ W^{\frac12}$, $g_- = (W^{\frac12} S^-)^{-1}$,
so $G^*$ is actually a $2^{\mathrm{rank}\  G}$-fold cover of the space of Stokes data, with the fiber given by the square roots of $W \in H$.}
Thus we have a map $\mu: \cO(G^*) \to \cA_1$.
Moreover, this map is actually a Poisson map, and it is known that it lifts to a map between the standard quantizations on the two sides: on the one hand $\cA_1$ quantizes to $\cA_\fq$ as we have been discussing, on the other hand $\cO(G^*)$ can be quantized to give the quantum group $U_q(\fg)$, and there is a canonical algebra map $U_q(\fg) \to \cA_\fq$ \cite{Gaiotto:2014lma,alex2019quantum,schradershapiro,shenqg}. 

If we take $\fq$ to be a primitive root of unity with
$\fq^{N^2} = 1$, 
then our general setup predicts that 
there must be a canonical multiwrapping map $\cA_{1} \to \cA_\fq$, whose image is central.
Anticipating a result from the next section, this multiwrapping map is in an appropriate sense local on $C$. As a consequence, we can focus just on the 
neighborhood of the irregular singularity.
Thus we predict that the composite map 
$\cO(G^*) \to \cA_1 \to \cA_\fq$ should factor through a map $\cO(G^*) \to U_q(\fg)$, in a manner independent of the geometry of the rest of $C$.
When $\fq^{N^2} = -1$ we have a similar statement,
replacing $G^*$ by its $\fq = -1$ analog which 
we call $G^*_-$.

In other words, our setup predicts that when $q$ is a primitive root of unity, the center of $U_q(\fg)$ 
should contain a copy of $\cO(G^*)$ or $\cO(G_-^*)$.
This is indeed a well known fact in the theory of quantum groups,
as described e.g. in \cite{MR1124981,BECK199429}.

Finally we remark that the map $U_q(\fg) \to \cA_\fq$ is compatible with the coproduct and R-matrix of $U_q(\fg)$, relating them to the amalgamation of character varieties
in the sense of \cite{MR2263192,alex2019quantum,shenqg}.
The amalgamation procedure can be given a physical interpretation, as governing the behaviour of the class $\cS$ theories in a limit where the gauge coupling associated to the irregular singularity becomes very large. It should be possible to use this interpretation to define 3d ${\cal N}=2$ interfaces which implement the amalgamation/coproduct, and thus physically 
derive the observation that when $q$ is a root of unity the coproduct 
localizes on its classical limit, i.e. the Poisson map  $G^* \times G^* \to G^*$ which makes $G^*$ into a Poisson-Lie group.

\subsubsection{Quantum groups and linear quivers}

Another physical setting where quantum groups appear is that of 4d linear quiver gauge theories with $n-1$ nodes and unitary gauge groups. In these theories, the 3d Coulomb branch can be described as a moduli space of solutions to the Bogomolny equations for $SL(n)$ on $\mathbb{R} \times \mathbb{C}^*$, in the presence of Dirac singularities. On general grounds, one can define an infinite collection of 
holomorphic functions on such a space (not linearly independent), organized as elements of a ``scattering matrix'' 
$T(z)$, a $n \times n$ matrix of Laurent polynomials. Remarkably, this structure survives quantization 
\cite{Bullimore:2015lsa,Frassek:2020lky}, 
in the form of an algebra map from the loop quantum group 
$U_q(L\mathfrak{sl}(n))$ to the quantized
Coulomb branch $\cA_\fq$.

This map can be given a physical interpretation via dualities starting from Witten's geometric engineering construction 
\cite{Witten:1997sc}. Witten's construction presents the 4d theory as sitting on a collection of D4 branes, coupled to a collection of NS5 branes. The worldvolume theory of the NS5 branes decouples in the 4d field theory limit. 
Nevertheless, the 4d theory must admit a supersymmetric coupling to the 6d worldvolume theory. 
Upon compactification on $M_q$, we expect the 6d theory to reduce to 4d holomorphic-topological Chern-Simons theory with coupling $(2 \pi \I)^{-1} \log q$ compactified on $\mathbb{C}^*$ \cite{Costello:2021zcl}, while the 4d theory reduces to a quantum-mechanical system coupled to the HT Chern-Simons theory as a topological line defect.\footnote{The 4d theory really reduces to a 2d system defined on an half-plane, but the 2d path integral effectively produces an ``analytically continued'' 1d system at the boundary \cite{Witten:2010zr}, with operator algebra $\cA_\fq$.}. This conjecture is a trigonometric version of the constructions in \cite{Ishtiaque:2018str,Costello:2021zcl,Dedushenko:2020yzd}. The coupling can be precisely expressed via Koszul duality as an algebra map from an ``universal gauge algebra'' $U_q(L\mathfrak{sl}(n))$ for the 4d CS theory to the algebra of local operators of the quantum-mechanical system, i.e. $\cA_\fq$.

The role of $M_q$ in this discussion strongly suggests that the root-of-unity behavior we have discussed should also apply to $U_q(L\fg)$ itself, i.e. $U_q(L\fg)$ should have a center isomorphic to the space of functions of scattering matrices $T(z)$ (or its $\fq=-1$ analog). This should be compatible with the root-of-unity phenomena in the individual 4d gauge theories. As $U_q(L\fg)$ and its scattering matrices play an important role in integrability, it would be nice to explore the potential applications of the root-of-unity construction in that context, e.g. for the Chiral Potts Model. 

\subsubsection{Combining these constructions}

In the last two subsections we discussed two different
ways of relating quantum groups to quantized Coulomb branches: either using class $\cS$ theories or linear quivers.
There are some examples which can be obtained in both ways. In these cases the expected map $U_q(L\fg) \to \cA_\fq$ associated to the brane realization factors through the well-known map $U_q(L\fg) \to U_q(\fg)$ to give the map $U_q(\fg) \to \cA_\fq$ arising from the class $\cS$ construction. 

The simplest example is a linear quiver with 
gauge groups $U(1) \times \cdots \times U(n-1)$ and $n$ flavours at the last node. This quiver theory admits a class $\cS$ realization with $\fg = \fsl(N)$, associated to a sphere 
with a rank $1$ irregular singularity and a regular singularity,
following a general recipe given in
\cite{Witten:1997sc,Gaiotto:2009hg}.

In this case the character variety is parameterized by $S^\pm$ and $W$ with the constraint that $S^+ W S^-$
belongs to a fixed conjugacy class, determined by the 
mass parameter for the flavour symmetry associated
to the regular singularity.
Thus the Coulomb branch is the intersection of $G^*$ and 
a regular orbit of $\fg$.
This Coulomb branch is quantized to a 
central quotient of $U_q(\fg)$.

Again, it should be possible to push our analysis to derive statements about the coproduct on $U_q(\fg)$ when $q$ is a root of unity.

\subsubsection{A simple example}

We continue with the setup of the last subsection, further specialized to the
case $n=2$ ($\fg = \fsl_2$). In this case the quiver theory is $U(1)$ gauge theory coupled to $2$ hypermultiplets of charges $\pm 1$, with a $SU(2)$ flavour symmetry acting on the hypermultiplets.
Thus this case is particularly simple since it only involves an abelian gauge group; in contrast,
for $n \ge 3$ we would have a nonabelian gauge theory.
A description of this example from the point of view of quantized cluster algebra 
is given in Example 4.7 of \cite{schradershapiro} and in Section 15.6 of \cite{alex2019quantum};
here we describe it from the point of view of the algebras of line operators.

Let 
$Z^{\pm 1}$ denote the Wilson lines with electric charges $\pm 1$, and let $X$, $Y$
be 't Hooft lines with magnetic charges $+1$ and $-1$ respectively. Let $\mu$ denote the
$SU(2)$ flavour mass.
In the quantized theory these line defects obey
\begin{align}
    X Y &= (1 + \fq^{-1} \mu Z)(1 + \fq\mu Z^{-1}) \, , \label{eq:qgrel-1} \\
    Y X &= (1 + \fq \mu Z)(1 + \fq^{-1} \mu Z^{-1}) \, , \\
    Y Z &= \fq^2 Z Y \, , \\
    X Z &= \fq^{-2} Z X \, . \label{eq:qgrel-4}
\end{align}
These relations are similar to those we discussed in \autoref{sec:u1-one-hyper}, and can similarly be derived from wall-crossing.
When $q^N = 1$,
the multiwrapping map $\cA_\pm \to \cA_\fq$ is given by $X_N = X^N$, $Y_N = Y^N$, $Z^{\pm 1}_N = Z^{\pm N}$, just as in \autoref{sec:u1-one-hyper}. 

The relations \eqref{eq:qgrel-1}-\eqref{eq:qgrel-4} imply in particular
\begin{align}
    XY - YX &=  (\fq^{-1}-\fq) \mu(Z- Z^{-1}) \, , \\
    XY + YX &- (\fq^{-1}+\fq) \mu(Z+ Z^{-1}) = 2 + 2 \mu^2 \, .
\end{align}
Then if we introduce
\begin{align}
    E &= \frac{X}{\fq^{-1}-\fq} \, ,  \\
    F &= \mu^{-1} \frac{Y}{\fq^{-1}-\fq} \, ,  \\
    K &= Z \, ,
\end{align}
we see that $E$, $F$, $K$ satisfy the standard defining relations of $U_q(\fsl_2)$,
with the quadratic Casimir $\frac12((\fq^{-1}-\fq)^2(EF+FE)-(\fq^{-1}+\fq)(K+K^{-1}))$ fixed to 
$\mu + \mu^{-1}$.

Setting $\fq = -1$ in the relations \eqref{eq:qgrel-1}-\eqref{eq:qgrel-4} 
we see that $\cA_{-}$ is 
a commutative algebra on generators $X$, $Y$, $Z^{\pm 1}$, $\mu^{\pm 1}$
with the relation $\mu^{-1} XY = \mu + \mu^{-1} - (Z + Z^{-1})$.
This algebra is naturally identified with $\cO(G^*_-) \simeq \cO(G^*)$.
Indeed, $Z^{\pm 1}$ become the eigenvalues of the formal monodromy $W$; $X$ and $\mu^{-1} Y$ become the off-diagonal entries of the Stokes matrices 
$S^+ W^{1/2}$ and $W^{1/2} S^-$; 
$\mu^{\pm 1}$ become the eigenvalues of the actual monodromy $S^+ W S^-$.
This version of the $q \to 1$ limit, giving a commutative algebra, is the one that arises naturally in our 4-dimensional setup.

In contrast, the more conventional description of the 
$q \to 1$ limit of $U_q(\fg)$ leads to the noncommutative algebra $U(\fg)$.
That limit corresponds to
reducing our 4-dimensional setup to a 3-dimensional theory in $\Omega$-background, by
writing $q = \e^{2 \pi \I R \hbar}$ and taking $R \to 0$ at fixed $\hbar$, as we discussed
in \autoref{sec:bulklines}.

We can also give an explicit geometric representation for the Wilson and 't Hooft lines in this theory, using its class $\cS$ description as the compactification of the 6d $(2,0)$ theory of type $\fsl(2)$ on a sphere with a regular singularity and a rank-$1$ irregular singularity. This is illustrated in \autoref{fig:qgUV}, where we cut out an infinitesimal disc around the irregular singularity and draw its complement in the sphere. There are two Stokes directions and two anti-Stokes directions associated with the irregular singularity. The Wilson lines $Z^{\pm 1}$ with electric charges $\pm 1$ are represented by loops traveling near the boundary of the disc. The 't Hooft lines $X$, $Y$ with magnetic charges $\pm 1$ correspond to open paths beginning and ending at the anti-Stokes directions. 
As in \autoref{sec:u1-one-hyper} these paths carry some additional sheet labels.
We note that the flavor charges of the 't Hooft lines realized here are slightly different from those of \eqref{eq:qgrel-1}-\eqref{eq:qgrel-4}: the blue open paths in \autoref{fig:qgUV} correspond to lines $X', Y'$ with
\begin{equation}\label{eqn:X'Y'}
 X' = X, ~ Y' = \mu^{-1} Y.   
\end{equation}

\insfigsvg{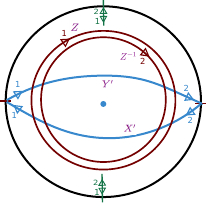}{1.5}{Here we illustrate $\mathbb{CP}^1$ minus an infinitesimal disc around an irregular singularity placed at infinity. There are two Stokes directions at the boundary of the disc, represented by green markers, and two anti-Stokes directions represented by red markers. The 't Hooft lines $X'$ and $Y'$ (in blue) correspond to open paths ending on the red markers at the boundary. The Wilson lines $Z$ and $Z^{-1}$ correspond to loops traveling around the boundary.}

From the geometrical point of view, the relations \eqref{eq:qgrel-1}-\eqref{eq:qgrel-4} obeyed by line defects follow from $\fsl(2)$ skein relations
on the punctured disc.\footnote{Similar relations between quantum groups and stated skein algebras appear in \cite{MR4493620}.}
As we did in \autoref{sec:u1-one-hyper} we check
these relations by lifting to the abelian skein algebra associated with a branched double cover of the disc, following the prescriptions of \cite{Neitzke:2020jik}. For example, \autoref{fig:qglifts} illustrates four lifts of $X'Y'$ into the abelian skein algebra, which yields
\begin{equation}
X'Y'=\mu + \mu^{-1} + \fq^{-1} Z + \fq Z^{-1}.
\end{equation}
This recovers \eqref{eq:qgrel-1} up to the convention \eqref{eqn:X'Y'}. 

\insfigsvg{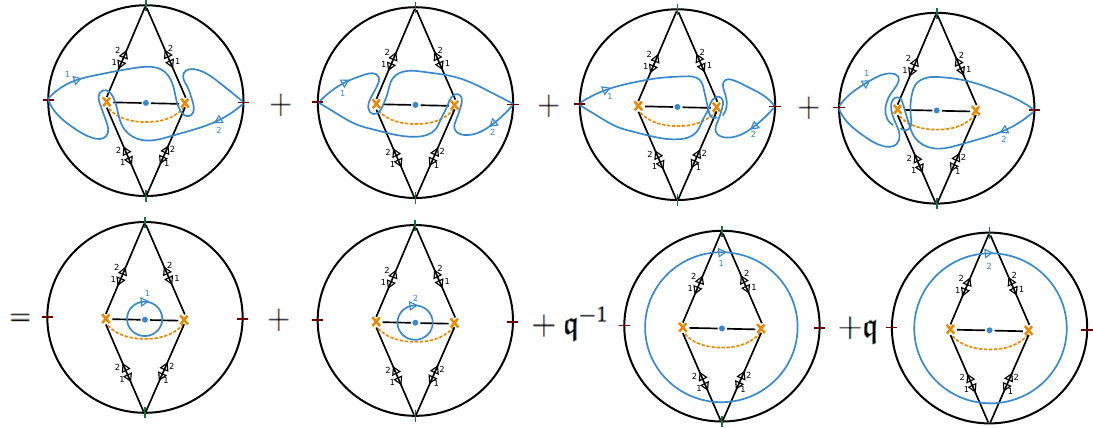}{0.9}{Here we illustrate the four lifts of $X'Y'$ in the abelian skein algebra, together with their simplification using the abelian skein relations. }

\section{Class \texorpdfstring{$\cS$}{S} theories and the Kapustin-Witten twist}

There is an alternative point of view on the
root-of-unity phenomena in class $\cS$ theories, which emerges from a connection to the 
Kapustin-Witten twist of $N=4$ super Yang-Mills.

\subsection{The large center from Kapustin-Witten twist} \label{sec:center-from-KW}

Fix a surface $C$ and Lie algebra $\fg$. As in the previous section, we will ignore topological subtleties associated to the global form of the group. The $\fg$-character variety of $C$ coincides with the phase space of 3d Chern-Simons theory compactified on $C$. The natural classical observables on this phase space are (networks of) Wilson lines, or ``skeins,'' 
in $M = C \times \mathbb{R}$. When $\fg$ is a complex Lie algebra their algebra reproduces the algebra of holomorphic functions on the character variety. 

The Chern-Simons path integral provides a natural quantization of this classical phase space. The standard path-integration contour is only well-defined if the Chern-Simons level $\kappa$ is appropriately quantized. More generally, the path integral can be done in an ``analytically continued'' manner for any level 
$\kappa \in \C$ \cite{Witten:2010cx,Witten:2010zr}.  This theory provides a canonical quantization of the algebra of observables, i.e. a quantization of the algebra of holomorphic functions on the character variety; this is the skein algebra  $\mathrm{Sk}_\fq[\fg,C]$ with parameter $\fq = \exp \frac{\I \pi}{\kappa}$. 

The analytically continued Chern-Simons path integral on a 3-manifold $M$ can be given a more robust definition, as the path integral for the Kapustin-Witten twist \cite{Kapustin:2006pk} of 4d $\cN=4$ super Yang-Mills with parameter $\Psi = \kappa$,
on $M \times [0,\infty)_t$, with Neumann boundary conditions at $t=0$.
Thus the Chern-Simons theory is ``relative'' and cannot be assigned a canonical Hilbert space.

From this point of view, the skein algebra is realized by skeins of boundary Wilson line defects,
and the root-of-unity phenomena arise from a special property of the 
Kapustin-Witten twist, as follows. For a generic value of $\Psi$, 
the twisted theory does not admit any line defects. When $\Psi = p/q \in {\mathbb Q}$, the situation is
different: a 't Hooft-Wilson line labeled by electric and magnetic weights $(\lambda,\lambda^\vee)$ 
survives into the twisted theory exactly if $q \lambda = p \lambda^\vee$ \cite{Kapustin:2006pk}.
Thus for $\Psi \in {\mathbb Q}$ (equivalently, for $\fq$ a root of unity)
the twisted theory has a family of bulk line defects labeled by weights; they are 
pure Wilson lines in the special case $\Psi = \infty$, and pure 't Hooft lines
if $\Psi = 0$.
When bulk lines do exist, we can make skeins out of them, and then bring them
to the Neumann boundary; the resulting boundary skeins are central in the skein 
algebra $\Sk_\fq[\fg,C]$, since
they are free to move in the fourth dimension. Thus we obtain a large center 
in $\Sk_\fq[\fg,C]$, isomorphic to the algebra of functions on the character variety.

\subsection{A duality}

We have now given two different physical points of view on the appearance of the
large center in $\Sk_\fq[\fg,C]$: one in \autoref{sec:general-N2-centers} using an $\cN=2$ theory
of class $\cS$, the other in \autoref{sec:center-from-KW} using $\cN=4$ super Yang-Mills.

The two pictures are related by a chain 
of dualities.
Indeed, we simply need to consider the  6d $(2,0)$ theory of type $\fg$ on $M_q \times C \times \R$, with a twist along $C$.
We can view this setup in two ways:

\begin{itemize}
    \item Reducing first on the $C$ factor gives a class $\cS$ theory 
on $M_q \times \R$. This is the setup
we used in \autoref{sec:general-N2-centers}.

\item 
If we replace the $z$-plane in $M_q$ by a cigar and reduce along its circle direction, the 6d theory 
reduces to 5d super Yang-Mills theory on a half-space. The boundary corresponds to the locus
$z=0$ in $M_q$, and carries a principal Nahm pole boundary condition from the 5d perspective. Reducing along the $x$-circle in $M_q$ 
corresponds to a further circle reduction of the 5d super Yang-Mills, with a twist by the instanton charge; the resulting system is 4d $\cN=4$ SYM on $[0,\infty) \times C \times \R$, 
with coupling $\Psi = \frac{1}{\pi \I} \log \fq$ and a principal Nahm pole boundary condition. Finally, S-duality maps this to coupling $\Psi^{-1} = \frac{1}{\pi \I} \log \fq$ and a Neumann boundary condition. This is the setup we discussed in  
\autoref{sec:center-from-KW}.

\end{itemize}

This sketchy derivation can be made more precise by twisting. The 6d $(2,0)$ theory admits a holomorphic-topological twist with four topological directions.
We can repeat the above reduction using the twisted theory 
on $M_q \times C \times \R$, where the holomorphic direction is the $z$-plane in $M_q$.
The twisted theory then reduces to
the holomorphic-topological twist of the class $\cS$ theory on $M_q \times \R$, and to the
Kapustin-Witten twist of $\cN=4$ super Yang-Mills on $[0,\infty) \times C \times \R$.

Now let us revisit the skeins of Chern-Simons theory 
in $C \times \R$. From the 6-dimensional perspective, we can realize
these skeins as 2-dimensional defects: they are the product of a skein in $C \times \R$ and 
the geodesic $\psi_0$ in $M_q$. 
(Note that these defects continue to exist in the 6-dimensional holomorphic-topological twist, 
since they are extended only in the 4 topological directions.)
In the class $\cS$ theory, these defects 
reduce to supersymmetric line defects in $M_q \times \R$
wrapping $\psi_0$ in $M_q$.
In $\cN=4$ SYM, they become boundary Wilson line skeins lying on the Neumann boundary condition. 

When $q$ is a root of unity, we can also consider ``multiwrapping'' these 2-dimensional defects,
by putting them on the product of a skein in $C \times \R$ and one of the
geodesics $\psi_{z_0}$ in $M_q$.
These multiwrapped defects reduce to the multiwrapped line defects in the class $\cS$ theory, and to the bulk skeins in the $\cN=4$ SYM theory. Thus the two ways we obtained the large center
in $\Sk_\fq[\fg,C]$ are
indeed related by our 6-dimensional setup.

We conclude with a few remarks:

\begin{enumerate}
\item The 4d SYM perspective allows one further manipulation. We can find 
an element of the $SL(2,\Z)$ duality group (generalized $S$-duality) which maps $\Psi = m/N$ to $\Psi = \infty$, where the bulk lines are Wilson lines. Correspondingly, this duality maps the Nahm boundary condition to a more general ``$(N,m)$-fivebrane'' boundary condition. 

We can attempt to describe this $(N,m)$-fivebrane boundary condition as the result of coupling the 4d gauge theory to a 3d SCFT ${\cal T}_{N,m}[\fg]$.
To do so, we consider
the 4d theory on an interval, where at one end we place the
$(N,m)$ boundary condition, and at the other end we place a 
Dirichlet boundary condition; the resulting 3d theory should be
${\cal T}_{N,m}[\fg]$. 
The Dirichlet boundary condition fixes the value of the bulk skeins, so effectively 
the algebra of skeins in ${\cal T}_{N,m}[\fg]$ is a quotient of the full skein algebra,
where the central elements are fixed to specific values determined by a point of the character
variety. 
In this way one should be able to identify the theories ${\cal T}_{N,m}[\fg]$ with the ones
considered in \cite{Creutzig:2021ext}.

\item A direct consequence of the 4d SYM 
description is that the construction of $L_N$ from $L$ 
must be a local operation on the skein representing $L$: a bulk line labelled by a representation $R$ must correspond to a boundary line in some (possibly virtual) representation $R_N$.\footnote{In principle this construction could also depend on the numerator $m$ in $\Psi = m/N$, through the choices of intertwiners at junctions.} We have already seen examples of this, for skeins consisting of a single loop. Similarly, the intertwiners at the junctions of bulk lines must map universally to intertwiners of boundary lines, not depending on the rest of the skein.

In other words, the construction of $L_N$ should follow from a categorical statement about the braided category of lines in analytically continued Chern-Simons theory and the center which appears at roots of unity.

\item Class $\cS$ theories come equipped with a large collection of ``class $R$'' boundary conditions, arising from the compactification of the $(2,0)$ 6d theories on 3d manifolds $D$ with cylindrical asymptotic regions \cite{Dimofte:2013lba}. 

In the 3d Coulomb branch perspective, such a boundary
condition is supported on the Lagrangian submanifold of flat connections on $C$ which extend to $D$. Boundary line defects arise from skeins in $D$; upon quantization they give rise to skein modules. For some background on the notion of skein module see e.g. \cite{MR1194712,2023arXiv230214734J}.

Following the general picture in \autoref{sec:line-ops-boundary}, we predict some 
additional structure when $q$ is a root of unity. 
First, the skein module associated to a class $R$ boundary condition 
will be supported on its classical image in the character variety.
Second, this skein module will admit a large submodule
which has a natural commutative algebra structure. 
Concretely, this algebra structure arises from the fact
that the skeins in the image of the multiwrapping map can be freely isotoped across one another,
so there is a well defined operation of juxtaposing them.
\end{enumerate}

\subsection*{Acknowledgements}
We thank Pablo Boixeda Alvarez, 
Francis Bonahon, Daniel Douglas, Alexander Goncharov, David Jordan, and Ivan Losev for useful discussions.
This research was supported in part by a grant from the Krembil Foundation. D.G. is supported by the NSERC Discovery Grant program and by the Perimeter Institute for Theoretical Physics. Research at Perimeter Institute is supported in part by the Government of Canada through the Department of Innovation, Science and Economic Development Canada and by the Province of Ontario through the Ministry of Colleges and Universities. 
A.N. is supported by NSF grant DMS-2005312.
G.M. and F.Y. is supported by DOE grant DE-SC0010008.

\newpage

\bibliographystyle{utphys}

\bibliography{SkeinCenter}

\end{document}